\documentclass[twocolumn,showpacs,preprintnumbers,amsmath,amssymb]{revtex4-1}

\usepackage{color}

\selectfont 

\usepackage{graphicx}       
\usepackage{dcolumn}        
\usepackage{bm}             
\usepackage{psfrag}

\newcommand{\nn}{\nonumber}

\newcommand{\lam}{L}

\newcommand{\gam}{\gamma}
\newcommand{\bet}{\beta}

\newcommand{\eps}{\epsilon}

\newcommand{\bx}{\mbox{\boldmath $x$}}

\newcommand{\B}{\mathcal{B}}
\newcommand{\Bef}{\B_{ln}}
\newcommand{\rstar}{r_\ast}

\newcommand{\uin}{u_{l\omega}^{\text{in}}}
\newcommand{\uup}{u_{l\omega}^{\text{up}}}

\newcommand{\beq}{\begin{equation}}
\newcommand{\eeq}{\end{equation}}

\newcommand{\Aout}{A^\text{(+)}_{l\omega}}
\newcommand{\Ain}{A^\text{(--)}_{l\omega}}

\newcommand{\Ainqn}{A^\text{(--)}_{l\omega_{ln}}}

\newcommand{\Gret}{G_{\text{ret}}} 
\newcommand{\Gtil}{\tilde{G}}

\newcommand{\Bout}{B_\text{out}}
\newcommand{\Bin}{B_\text{in}}
\newcommand{\Cin}{C_\text{in}}

\newcommand{\alp}{\alpha}

\newcommand{\GQNM}{\Gret^{\text{QNM}}}
\newcommand{\Gsing}{\Gret^{\text{sing.}}}

\begin{document}

\preprint{}

 \title{Wave Propagation and Quasinormal Mode Excitation\\ on Schwarzschild Spacetime}
 
\author{Sam R. Dolan}
 \email{s.dolan@soton.ac.uk}
 \affiliation{%
 School of Mathematics, University of Southampton, Highfield, Southampton SO17 1BJ, UK.
}%

\author{Adrian C. Ottewill}
 \email{adrian.ottewill@ucd.ie}
 \affiliation{
 Complex and Adaptive Systems Laboratory / School of Mathematical Sciences, University
College Dublin, Belfield, Dublin 4, Ireland
 }
\date{\today}

\begin{abstract}
To seek a deeper understanding of wave propagation on the Schwarzschild spacetime, we investigate the  relationship between (i) the lightcone of an event and its caustics (self-intersections), (ii) the large-$l$ asymptotics of quasinormal (QN) modes, and (iii) the singular structure of the retarded Green function (GF) for the scalar field. First, we recall that the GF has a (partial) representation as a sum over QN modes. Next, we extend a recently-developed expansion method to obtain asymptotic expressions for QN wavefunctions and their residues. We employ these asymptotics to show (approximately) that the QN mode sum is singular on the lightcone, and to obtain approximations for the GF which are valid close to the lightcone. These approximations confirm a little-known prediction: the singular part of the GF undergoes a transition each time the lightcone passes through a caustic, following a repeating four-fold sequence. We conclude with a discussion of implications and extensions of this work.
\end{abstract}

\maketitle

%
%

\section{Introduction}

Characteristic damped resonances of black holes known as \emph{quasinormal (QN) modes} have received much attention over the last three decades \cite{Vishveshwara, Press, Goebel, Chandrasekhar-Detweiler, Detweiler, Chandrasekhar, Ferrari-Mashhoon, Mashhoon, Schutz-Will-1985, Leaver-1985, Kokkotas, Nollert, Dolan-Ottewill, Cho-Cornell-Doukas-Naylor-2010, Dolan-2010}. QN modes are expected to play a key role in strong-field processes in black hole dynamics (see e.g.~\cite{Kokkotas-Schmidt, Berti-Cardoso-Will-2006, Ferrari-Gualtieri, Berti-Cardoso-Starinets}). For example, after the merger of two black holes, the composite system undergoes a `ringdown' phase during which it sheds its asymmetries through gravitational radiation. This phase is well-described in the linearized approximation as a black hole radiating via the least-damped, low-multipole QN modes. 
 
A black hole QN mode corresponds to a complex frequency, $\omega_{Q} = 2 \pi ( 1/T - i/ \tau)$, where $T$ is the period of oscillation and $\tau$ is the decay timescale. The frequencies of the QN spectrum are labelled by integers, e.g.~the multipole $l$ and overtone $n \ge 0$ (in Schwarzschild spacetime). The QN spectrum depends only on the properties of the field (e.g.~its spin $s$) and the black hole geometry (e.g. mass~$M$, angular momentum $J$), i.e.~$\omega_{ln}(s, J, M)$. Hence, the (future) observation of QN modes in gravitational-wave data would allow for the direct inference of mass and angular momentum of black holes \cite{Berti-Cardoso-Will-2006}.

Most studies of QN modes have focussed on the challenge of determining the QN frequency spectrum. Rather fewer studies have tackled the more demanding question of how (and how much) QN modes are excited by a given initial data (see notable exceptions \cite{Leaver-1986, Sun-Price, Andersson-1997, Nollert-Price, Berti-Cardoso-2006, Dorband}). Although the QN formalism is well understood \cite{Leaver-1986}, such calculations face certain technical challenges. For example, unlike \emph{normal modes} in conservative systems, QN modes do not form a complete set (i.e. they provide only a partial description of the system). Furthermore, QN mode sums may not be strictly convergent. The aim of this work is to show that, despite technical issues, the QN mode formalism can be used to gain geometric insight into intriguing features of wave propagation on curved spacetimes. In this study, we consider the simplest case: a non-rotating ($J=0$) black hole of mass $M$ in the Schwarzschild coordinate system $\{t,r,\theta,\phi\}$  described by line element
\beq
ds^2 = -f(r) dt^2 + f^{-1}(r) dr^2 + r^2 \left( d\theta^2 + \sin^2 \theta  d \phi^2 \right) ,
\eeq
where $f(r) = 1 - 2M/r$. The event horizon is at $r= 2M$.

\begin{figure*}
 \begin{center}
\begin{tabular}{cc}
  \includegraphics[width=8cm]{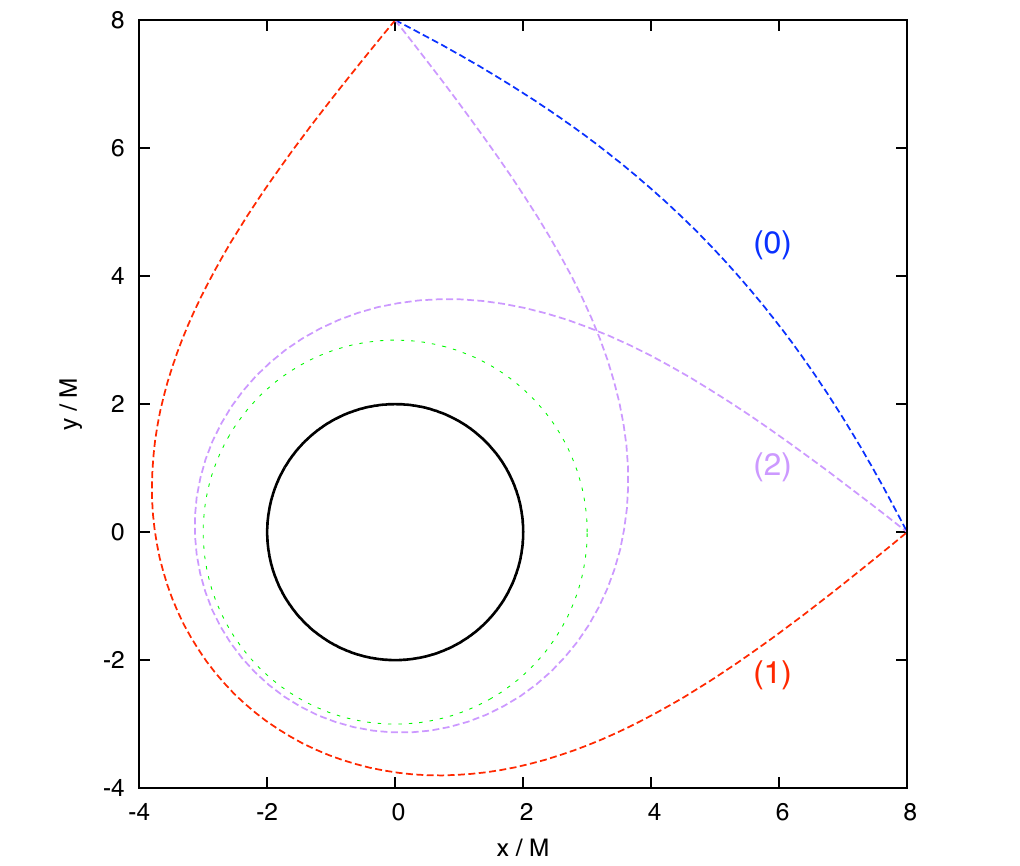}&
 \includegraphics[width=\columnwidth]{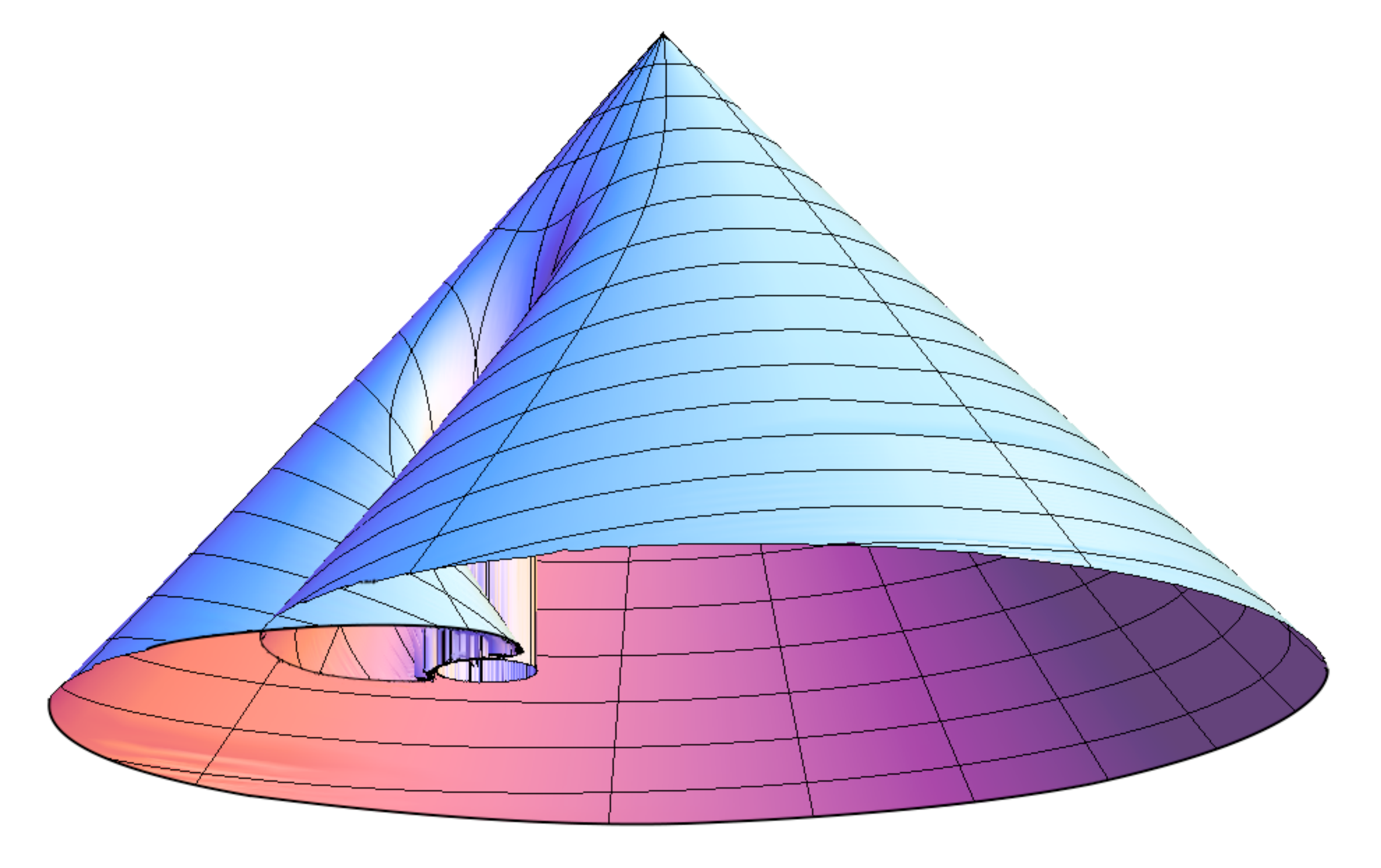}
\end{tabular}
 \caption{\emph{Left:} Light rays orbiting a Schwarzschild black hole. This plot shows that two points at fixed spatial coordinates in the vicinity of a black hole are linked by multiple null geodesics (i.e.~light rays). Here we show the `direct' ray (ray~0) [passing through $\Delta \phi = \pi/2$] and two orbiting rays [passing through $\Delta \phi = 3 \pi / 2$ (ray 1) and $\Delta \phi = 5 \pi / 2$ (ray 2)] in the $xy$ plane.  \emph{Right:} The self-intersecting light cone (reproduced from Fig.~1.1 of \cite{Wardell:thesis}; see also \cite{Perlick}). The right plot shows the lightcone of a spacetime point (at the apex of the cone) in the vicinity of a black hole (the horizon of the BH is visible as the smaller circle). The coordinate time $t$ runs vertically downwards, and one spatial dimension ($z$) has been suppressed. Note the formation of caustics (lines along which the lightcone intersects itself). See also Fig.~\ref{fig:snapshot} and \ref{fig:lightcone-gf}. }
 \label{fig:lightcone3d}
 \end{center}
\end{figure*}

To motivate this study, let us begin with a simple thought experiment, in which a `beacon' at spatial coordinate $\overrightarrow{x}'$, close to a black hole, emits an instantaneous flash of light at $t=0$. A moment later, at $t = \delta t$, the light forms an almost-spherical wavefront around $\overrightarrow{x}'$. As $t$ increases, the wavefront propagates outwards, describing a `lightcone': a 3D null hypersurface embedded within the 4D spacetime. Locally, each part of the wavefront propagates along a null geodesic of the spacetime. The presence of the black hole warps the wavefront, and the part of the wavefront that passes close to the `unstable photon orbit' at $r=3M$ may orbit the black hole multiple times. This observation has several concomitants: (i) the lightcone intersects itself along null lines in spacetime called \emph{caustics}; the first caustic forms after a time $\Delta t$, where $\Delta t$ is the shortest time it takes for a light-ray emanating from $\overrightarrow{x}'$ to pass to the diametrically opposed side of the black hole; (ii) the wavefront passes through the spatial point at $\overrightarrow{x}'$ (or indeed any point) an infinite number of times, at $t = t_1, t_2, \ldots$ (where $t_1 < t_2 < \ldots$), although each `echo' will be fainter than the last; and (iii) a black hole acts as a very strong gravitational lens, generating an infinite number of images of the same source \cite{Perlick, Bozza-2008, Bozza-Mancini-2009}. Some aspects of this thought-experiment are illustrated in Fig.~\ref{fig:lightcone3d}.

In this paper, we study the retarded Green function which provides a mathematical description of wave propagation. In flat spacetime, the retarded Green function (GF) has support only on the lightcone itself. In curved spacetimes, it is well-known that the GF also has (non-singular) support within the lightcone, due to backscattering from the curved geometry \cite{Hadamard, Friedlander}. It is perhaps less well-known that the singular part of the GF is modified when caustics are encountered \cite{Ori1, CDOW1, CDOW2}. The principal aim of this paper is to show that the singular part of the GF on the Schwarzschild spacetime may be well-understood through the large-$l$ asymptotics of the QN mode representation. 

In Ref.~\cite{Dolan-Ottewill} (henceforth Paper I) we developed an expansion method to investigate the large-$l$ asymptotics of the QN frequency spectrum of spherically-symmetric black holes. We showed that the frequencies of the lightly-damped modes of the Schwarzschild black hole are given by the expansion
\begin{align}
&\omega_{ln} \approx \frac{1}{\sqrt{27}M} \left( L - i N + \left[ \frac{\beta}{3} - \frac{5N^2}{36} - \frac{115}{432} \right] L^{-1} \right.\nonumber \\
& \quad\left. - i N \left[ \frac{\beta}{9} + \frac{235N^2}{3888} - \frac{1415}{15552} \right] L^{-2} + \mathcal{O}(L^{-3})  \right),
\label{freq-expansion}
\end{align}
where $\beta = 1-s^2$ and
\beq
L = l + 1/2 , \quad \quad N = n + 1/2,  \label{eq-LN}
\eeq
and $l \ge |s|$ and $n \ge 0$ are the angular momentum and overtone numbers, respectively.
Higher-order terms in the Schwarzschild expansion were given in Paper I. Expansions for some other spacetimes were computed in \cite{DOC1, Dolan-2011, DOC2}.

In this work, we develop the expansion method further to investigate the asymptotics of the so-called `excitation factors' $\Bef$ (defined in Sec.~\ref{sec:gf}), which provide a stepping-stone to the properties of the Green function. 

The remainder of this paper is organised as follows. In Sec.~\ref{sec:basics} we outline the basics of black hole perturbation theory and quasinormal modes. In Sec.~\ref{sec:gf} we recap the theory that leads to a (partial) decomposition of the Green function in a series of QN modes. In Sec.~\ref{sec:asymptotics} we derive some key results for the large-$l$ asymptotics of QN wavefunctions and excitation factors. In Sec.~\ref{sec:structure} we employ the asymptotics to investigate the singular structure of the Green function in the Schwarzschild spacetime. We conclude in Sec.~\ref{sec:conclusion} with a discussion of the applications of the method. 

\section{Basics\label{sec:basics}}
The evolution of scalar-field, electromagnetic and gravitational perturbations on the Schwarzschild spacetime are governed by wave equations \cite{ReggeWheeler, Price}. The simplest example is the (minimally-coupled) scalar field in the vacuum exterior of a black hole, which satisfies a curved-space Klein-Gordon equation:
\beq
\Box \Phi = \frac{1}{\sqrt{-g}} \partial_{\mu} \left(\sqrt{-g} g^{\mu \nu} \partial_{\nu} \Phi \right) = 0 .
\eeq
Here $g = -r^4 \sin^2\theta$ is the metric determinant, and $g^{\mu \nu}$ is the contravariant version of the metric. The symmetry of the spacetime allows for a complete separation of variables, i.e.
\beq
\Phi(t,r,\theta,\phi) = \frac{1}{2\pi} \sum_{lm} Y_{lm}(\theta, \phi) c_{lm} \int u_{l \omega}(r) e^{-i \omega t} d\omega , 
\eeq
where $Y_{lm}(\theta,\phi)$ are spherical harmonics. The radial function $u_{l \omega}(r)$ satisfies the second-order ordinary differential equation
\beq
\left[ \frac{d^2}{d \rstar^2} + \omega^2 - f(r) \left( \frac{l (l+1)}{r^2} + \frac{2M\beta}{r^3} \right) \right] u_{l\omega}(r) = 0,   \label{rad-eq-1}
\eeq
where, as above, 
$
f(r) = 1-2M/r $,
 and, in addition, we have introduced  the standard `tortoise coordinate' $\rstar = r + 2M  \ln ( r/2M - 1)$. Here $\beta = 1-s^2$ with $|s| = 0, 1, 2$ for scalar, electromagnetic and gravitational perturbations \cite{Chandrasekhar}, respectively. On a black-hole spacetime, physically valid solutions are purely `ingoing' at the event horizon. Let us define `ingoing' solutions $\uin(r)$ via
\beq
\uin(r) \sim 
\left\{ \begin{array}{ll} e^{-i \omega \rstar}, &  \rstar \rightarrow -\infty ,  \\ 
  \Ain e^{-i \omega \rstar} + \Aout e^{i \omega \rstar},   &    \rstar \rightarrow +\infty . \end{array} \right.
  \label{uin-def}
\eeq
where $\Ain$ and $\Aout$ are complex constants. Quasinormal modes (QNM)  are modes with complex frequencies $\omega_{ln}$ such that $\Ainqn = 0$ (where $n = 0, 1, \ldots$ is the overtone number). In other words, QNMs are purely ingoing at the horizon and purely outgoing at spatial infinity.

\begin{figure}
 \begin{center}
 \includegraphics[width=\columnwidth]{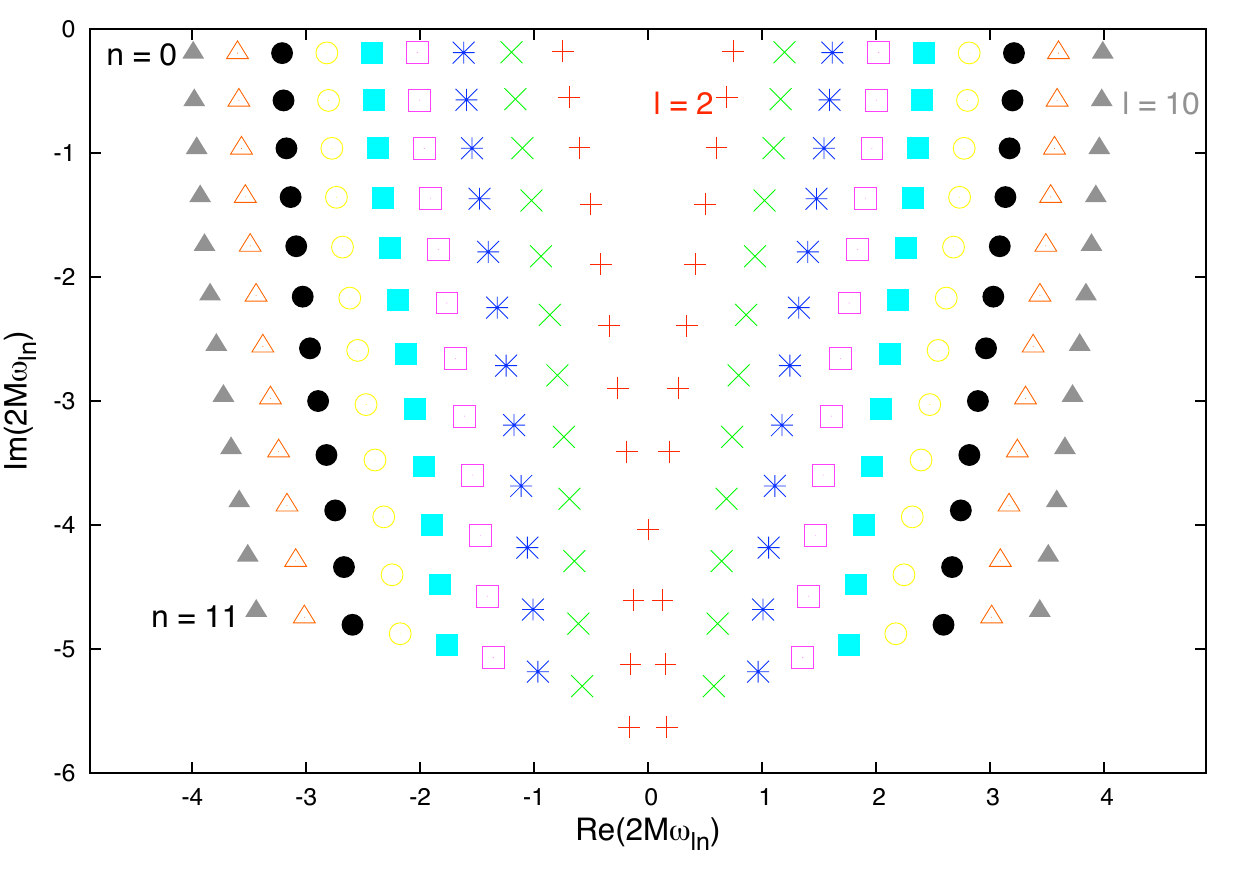}
 \caption{\emph{The gravitational quasinormal mode spectrum of the Schwarzschild black hole}. The plot shows the real and imaginary parts of the gravitational ($|s| = 2$) QNM frequencies, for angular momenta $l = 2 \ldots 10$ and overtones $n = 0 \ldots 11$.}
 \label{fig:freqs}
 \end{center}
\end{figure}

The spectrum of QNMs of the Schwarzschild spacetime is illustrated in Fig.~\ref{fig:freqs}. The plot highlights some key properties. Firstly, all QNM frequencies have a negative imaginary part (in other words, the amplitude of all QNMs decay with time). Secondly, the frequencies are approximately evenly-spaced in $l$ and $n$. In the limit $n\rightarrow \infty$ (for fixed $l$) the spacing $\omega_{l,n+1} - \omega_{l,n}$ approaches an imaginary (and $l$-independent) constant. 
In the limit $l \rightarrow \infty$ (for fixed $n$) it is well-established \cite{Goebel} that the spectrum has the asymptotic form
$
\omega_{ln} \approx  L  \Omega - i N \Lambda , \label{eq:freq0}
$
where $L$ and $N$ are defined in Eq.~(\ref{eq-LN}), while $\Omega = (\sqrt{27} M)^{-1}$ is the orbital frequency on the `photon sphere' at $r = 3M$, and $\Lambda = (\sqrt{27} M)^{-1}$ is the so-called Lyapunov exponent (i.e.~decay timescale) \cite{Lyapunov1} for orbits in the vicinity of this sphere. In essence, the asymptotic properties of the QNM spectrum are linked to the properties of the photon sphere; this observation extends also to more general black hole spacetimes \cite{Dolan-2010}. In Paper~I, we developed an asymptotic expansion method to explore this link mathematically, and applied the method to obtain high-order expansions in inverse powers of $L$ for the low-overtone modes of the spectrum, e.g.~Eq.~(\ref{freq-expansion}). 
In this work, we will extend the method to investigate wave propagation in the large-$L$ limit.

\section{The Green function and quasinormal mode sums\label{sec:gf}}

The role of QNMs in wave propagation may be appreciated by examining the contribution of QNMs to the retarded Green function \cite{Leaver-1986, Andersson-1997}. Here, for simplicity, we will consider the retarded Green function for the scalar field on Schwarzschild spacetime, $\Gret(x,x')$, which is defined by 
\beq
\Box_x \Gret(x, x') = \delta^4(x - x')
\eeq
(where $\delta^4$ is a covariant version of the Dirac delta distribution) 
with appropriate causal conditions. The Green function may be used to solve two types of problem. Firstly, if a field $\Phi(x)$ is generated by a source $S(x)$ of compact support, i.e.~$\Box \Phi(x) = S(x)$, then the field is found from the spacetime integral
$
\Phi(x)  =  \int \Gret(x, x') S(x') d^4 x' . 
$
Secondly, if an unsourced field satisfies a Cauchy initial value problem then the field may be found via an integral over a hypersurface. For example, if at $t=0$ the field (in the exterior region of the BH) is given by $\Phi|_{t=0} = \Phi_0(\bx)$ and $\partial_t \Phi|_{t=0} = \dot{\Phi}_0$, then the field at later times $t$ is given by 
\begin{align}
\Phi(t,\bx) &= \int \left[ \Gret(t,\bx ; 0, \bx') \dot{\Phi}_0(\bx') +\right.\nonumber\\
&\left.  \partial_t \Gret(t,\bx; 0, \bx') \Phi_0(\bx') \right] d^3 \bx' .
\label{eq:evolution}
\end{align}

The retarded Green function on Schwarzschild spacetime may be expressed in terms of an inverse Laplace transform, 
\begin{align}
\Gret(x, x') = &\frac{1}{2 \pi r r'} \sum_{l=0}^\infty (2l+1) P_l(\cos \gamma)\times
\nonumber\\  &\int_{-\infty + ic}^{+\infty + ic} \Gtil_{l \omega}(r, r') e^{- i\omega(t-t')}  d \omega .  \label{Gret-mode-sum}
\end{align}
Here $x,x'$ are spacetime points at radii $r$, $r'$, separated by coordinate time $t - t'$ and spatial angle $\gamma$, [where $\cos \gam = \cos \theta \cos \theta' + \sin \theta \sin \theta' \cos(\phi - \phi')$] and $c$ is a positive constant, and $\Gtil_{l\omega}(r,r')$ is the `radial' Green function satisfying
\begin{align}
&\left[ \frac{d^2}{d \rstar^2} + \omega^2 - f(r) \left( \frac{(L^2-1/4)}{r^2} + \frac{2M}{r^3} \right) \right] 
\Gtil_{l\omega}(r,r') \nonumber\\ &\qquad=  -\delta(r_\ast - r_\ast^\prime) ,
\end{align}
together with ingoing (outgoing) boundary conditions at the horizon (spatial infinity) and a causality condition. The radial Green function may be constructed from a pair of linearly-independent solutions ($\uin$ [defined in Eq.~(\ref{uin-def})] and $\uup$ [defined by $\uup \sim e^{+i\omega \rstar}$ as $r_\ast \rightarrow \infty$]) of the homogeneous equation (\ref{rad-eq-1}), via
\beq
\Gtil_{l\omega}(r,r')  =  -\frac{\uin( r_{\ast<}) \, \uup( r_{\ast >}) }{W(l,\omega)}  ,
\eeq
where $r_{\ast>} = \max( r_\ast, r_\ast^\prime )$, $r_{\ast <} = \min( r_\ast, r_\ast^\prime )$ and the Wronskian $W(l,\omega)$ is
\beq
W(l, \omega) = \uin \frac{d \uup}{d r_\ast} - \uup \frac{d \uin}{d r_\ast} .
\eeq
Note that the Wronskian is zero at QNM frequencies, i.e.~$W(l, \omega_{ln}) = 0$. Equivalently, the `in' and `up' QNM solutions are linearly-dependent. Hence the QNM frequencies correspond to poles of $\Gtil_{l \omega}$. 

\begin{figure}
\begin{center}
\includegraphics[width=\columnwidth]{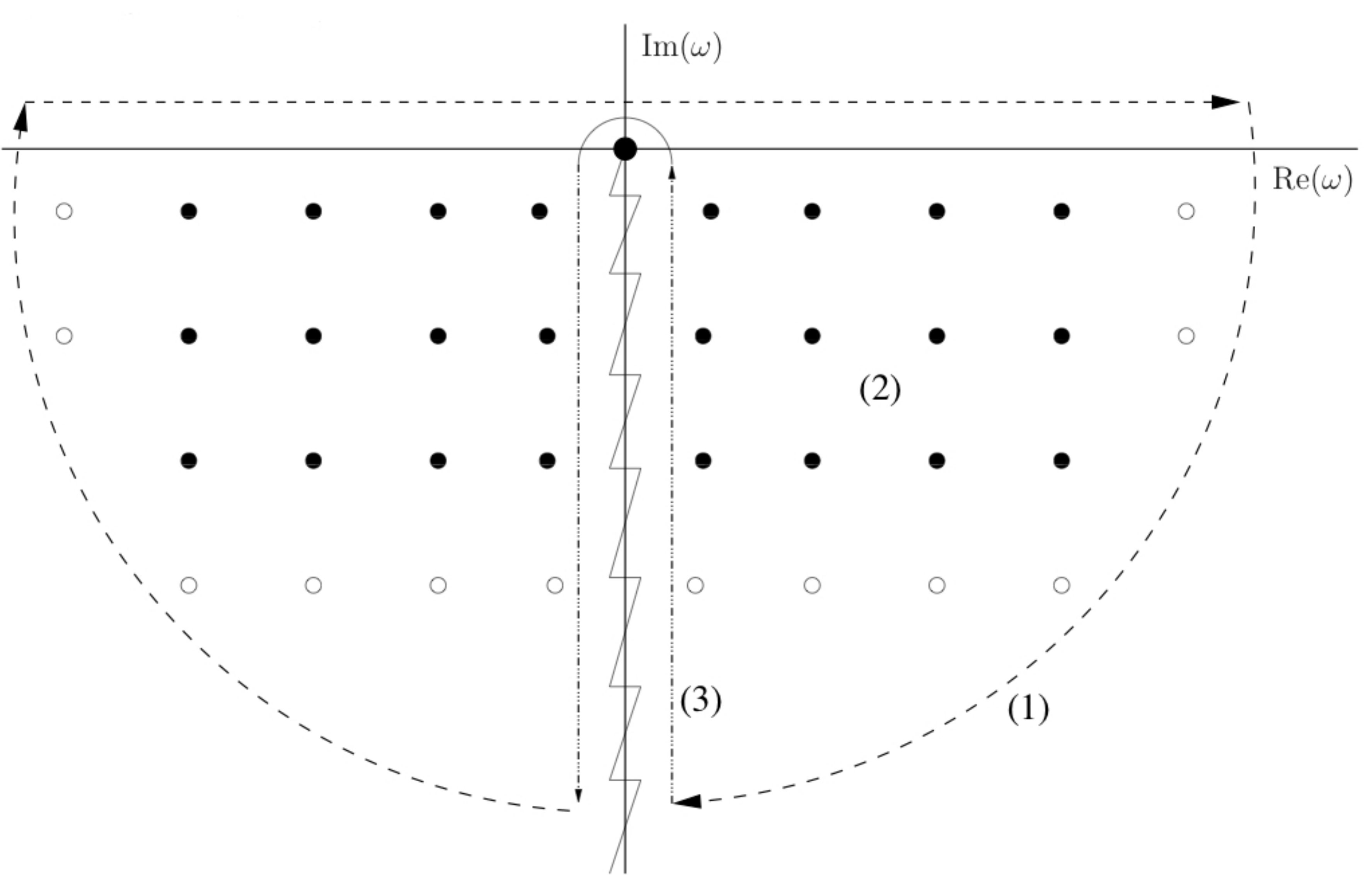}
\end{center}
\caption[]{\emph{Integration contour in the complex frequency plane}. In Eq.~(\ref{Gret-mode-sum}), the GF is expressed as an integral in the upper half of the complex-frequency plane. In principle, for $T > 0$ [Eq.~\ref{reflection-time}], this may be computed by closing the contour in the lower half-plane. This gives three contributions: (1) a high-frequency arc, (2) a sum over the residues of the QN modes (which are poles of the GF, indicated by the dots above), and (3) an integral along the branch cut which runs along the negative imaginary-frequency axis.}
\label{fig-qnm-contour}
\end{figure}

At suitably `late' times $t - t' > r_\ast + r_\ast^\prime$, the contour of integration defining the inverse Laplace transform (\ref{Gret-mode-sum}) may be deformed in the lower half-plane, as shown in Fig.~\ref{fig-qnm-contour}. The QNM contribution $\GQNM$ arises from applying Cauchy's integral theorem to pick up the contribution from the residues of the poles at QN frequencies $\{ \omega_{ln} \}$. It may be written in the following way:
\begin{align}
\GQNM = \frac{2}{r r'} \text{Re} \sum_{l=0}^\infty &\sum_{n=0}^\infty \left( 2 l+1 \right) P_l(\cos \gam)\times\nonumber\\
& \mathcal{B}_{ln} \tilde{u}_{ln}(r)\tilde{u}_{ln}(r') e^{-i \omega_{ln} T}  .  \label{G-QNM-def}
\end{align}
Here the `reflection time $T$' is defined by 
\beq
T \equiv t - t^\prime - r_\ast - r_\ast^\prime ,  \label{reflection-time}
\eeq
the normalised wavefunctions $\tilde{u}_{ln}(r)$ are defined by
\beq
\tilde{u}_{ln}(r) \equiv  u^{\text{in}}_{l \omega_{ln}}(r) \times \left[ \Aout e^{i \omega_{ln} \rstar} \right]^{-1},
 \label{u-norm}
\eeq
and the so-called `excitation factors' $\B_{ln}$ \cite{Leaver-1986} are defined by
\beq
\B_{ln} \equiv \left[  \frac{\Aout}{2 \omega} \left(  \frac{\partial \Ain}{\partial \omega}  \right)^{-1}  \right]_{\omega=\omega_{ln}}  .  \label{Bef-def}
\eeq

The `QNM sum' expression, given by Eq.~(\ref{G-QNM-def}) and arising from the residues of the poles (2) in Fig.~\ref{fig-qnm-contour}, was considered in some detail by Leaver \cite{Leaver-1986}. It is not the only contribution to the Green function; observe from Fig.~\ref{fig-qnm-contour} that we must also take into account the parts of the contour (1), the `high-frequency arcs', and (3), the `branch-cut integral'. Parts (1) and (3) are associated with the `prompt response' (direct propagation) contribution, and `power-law tail' (back-scattering from potential) contributions to the Green function \cite{Price, Ching}, whereas part (2) is associated with `quasinormal ringing' related to the peak of the potential barrier, which is associated with the photon sphere.

The QNM sum Eq.~(\ref{G-QNM-def}) seems ill-defined at `early' times $T < 0$, due to the exponential divergence of the $\exp(-i \omega_{ln} T)$ factors in the sum over overtones $n$ in Eq.~(\ref{G-QNM-def}). By contrast, at late times $T > 0$, it appears that the sum over $n$ will converge, due to the exponential suppression of high overtones. Andersson \cite{Andersson-1997} demonstrated the convergence of the sum over $n$ for fixed $r$. Note however, that if the source does not have compact support this does not imply convergence of the integral in Eq.~(\ref{eq:evolution}), since for any $T$  there will always be points in the integration outside the radius of convergence of Eq.~(\ref{G-QNM-def}). Furthermore, Andersson showed that, subject to the above caveat, a combination of the QNMs and the contribution from the branch cut can form a \emph{complete} description of the wave evolution for $T > 0$ (in other words, the prompt response [contour (1)] can be neglected in the regime $T > 0$).  

In principle, if we know the excitation factors $\B_{ln}$, frequencies $\omega_{ln}$ and wavefunctions $\tilde{u}_{ln}(r)$ we can deduce from (\ref{G-QNM-def}) the QNM response excited by a given perturbation. However, such calculations are relatively technical and have only been attempted in only a handful of studies in the literature \cite{Leaver-1986, Sun-Price, Nollert-Price, Andersson-1997, Berti-Cardoso-2006}. In this paper, we develop approximation methods in an attempt to investigate a physical question: what is the structure of the Green function close to the lightcone?

\section{Large-$l$ asymptotics of QN modes\label{sec:asymptotics}}

To analyse the large-$l$ asymptotics of the QNM sum Eq.~(\ref{Gret-mode-sum}), we need asymptotic approximations for (i)~the QN frequencies, (ii)~the QN wavefunctions and (iii)~the QN excitation factors (QNEFs). These ingredients are obtained in this section. We set $M=1$ in much of the following. 

\subsection{The expansion method}
In Paper I, we described the key steps in the expansion method. The first step is to introduce an ansatz for the wavefunction, 
\beq
u_{l\omega}(r) = \exp\left[\!i \omega \!\int^{r_\ast} \!\! \left( 1 + \frac{6}{r^\prime} \right)^{\!\!1/2} \!\left(1-\frac{3}{r^\prime}\right)dr_\ast^\prime  \!\right] \! v_{l \omega}(r) .  \label{ansatz}
\eeq
Note that the ansatz has a geometric motivation; the phase factor in Eq.~(\ref{ansatz}) closely resembles the (square root of the) right-hand side of the geodesic equation in the critical case $b = b_c = \sqrt{27}$ (where $b = l / \omega$ is the impact parameter for a geodesic approaching from spatial infinity), i.e.
\beq
\frac{b_c^2}{r^4} \left(\frac{dr}{d \gam} \right)^2 = \left(1 + \frac{6}{r} \right) \left(1 - \frac{3}{r} \right)^2 .
\eeq
Note also that the QN boundary conditions are automatically satisfied if $v_{l\omega}(r)$ is regular in both limits $\rstar \rightarrow \pm \infty$. Now, substitution of ansatz (\ref{ansatz}) into Eq.~(\ref{rad-eq-1}) leads to the equation
\begin{widetext}
\begin{eqnarray}
(f v')' + \left[ 2i\omega \left( 1+ \frac{6}{r} \right)^{1/2} \left(1 - \frac{3}{r} \right) \right] v'   \nn
+ 
 \left[ \frac{27 \omega^2 - \lam^2}{r^2} + \frac{27i \omega}{r^3} \left(1 + \frac{6}{r} \right)^{-1/2} + \frac{1}{4r^2} - \frac{2\beta}{r^3} \right] v &=& 0,    \label{rad-eq-2}
\end{eqnarray}
where here $'$ denotes differentiation with respect to $r$. 
The equation above looks more complicated than Eq.~(\ref{rad-eq-1}), but it has the advantage of being amenable to an expansion in inverse powers of $\lam$. We expand the frequency and the wavefunction in inverse powers of $L$ as follows:
\beq
\omega_{ln} = \sum_{j=-1}^{\infty} \varpi_{jn} L^{-k}   \label{omega-series}
\eeq
and
\beq
v_{ln}(r) = \left[ \left(1-\frac{3}{r}\right)^n + \sum_{i = 1}^{n} \sum_{j=1}^\infty a_{ijn} L^{-j} \left(1-\frac{3}{r}\right)^{n-i} \right] \exp \left( S_{0n}(r) + L^{-1} S_{1n}(r) + \ldots \right)
\label{v-expansion}
\eeq
\end{widetext}
Here $\varpi_{jn}$ and $a_{ijn}$ are complex coefficients, and $S_{jn}(r)$ are smooth functions of $r$. Inserting the ansatz into Eq.~(\ref{rad-eq-2}) and collecting together like-powers of $L$ leads to a system of equations. To solve the system, we impose a condition of regularity on the functions $S_{jn}$ at $r = 3M$. This leads directly to the frequency expansion Eq.~(\ref{freq-expansion}) and to the results quoted below. More detail on the expansion method is given in Paper~I, and in Appendix~\ref{appendix-Bef}.

\subsection{QN wavefunctions}
To leading order in $L$, the QN wavefunctions are
\begin{align}
 \tilde{u}_{ln}(r) &\approx C_{n} \exp({S_{0n}(r)})  (1-3/r)^n \nonumber
\\
&\exp \left(  i \omega_{ln} \int_3^r \left[ \left(1+6/r\right)^{1/2} \left(1 - 3/r \right) - 1  \right] f^{-1} dr \right) \,  \label{utilde-1}
\end{align}
where $C_{n}$ is a normalisation constant such that $\lim_{r \rightarrow \infty} \tilde{u}_{ln}(r)  =  1$. The function $S_{0n}(r)$ is found to be
\beq
\quad \quad S_{0n}(r) = \frac{1}{2} \ln ( 2 / x ) + 2 (n+1/2) \ln \left( \frac{2 + \sqrt{3}}{x + \sqrt{3}} \right)   \label{S0-def1}
\eeq
where
\beq
x \equiv \left( 1 + \frac{6}{r} \right)^{1/2}.  \label{x-def}
\eeq
We makes three notes in passing: (i)~$x$ varies between $1$ (at spatial infinity) and $2$ (at the horizon), (ii)~the constant of integration in finding $S_{0n}(r)$ was chosen such that $S_{0n}(r=2) = 0$; (iii)~to leading order the radial wavefunctions are independent of the spin of the field. 

The integral in Eq.~(\ref{utilde-1}) can be found in closed form,
\begin{align}
\exp &\left( \int_3^r \left[ (1+6/r)^{1/2} (1-3/r) - 1 \right] f^{-1} dr \right)
=\nonumber\\
&\frac{1}{4} (2-\sqrt{3})^6 \, e^{3-\sqrt{27}} e^{-r(1-x)} \left( \frac{1+x}{2-x} \right)^4 ,
\end{align}
hence
\begin{align}
\tilde{u}_{ln}(r) \approx  \left(1 - 3/r \right)^n &x^{-1/2} e^{-i \omega \frac{1}{2}r(x-1)^2} \nonumber\\
&\left( \frac{1+x}{4-2x}  \right)^{4 i \omega} \left( \frac{1 + \sqrt{3}}{x + \sqrt{3}} \right)^{2n+1} 
\end{align}
Inserting the lowest-order approximation for the frequency, $\omega_{ln} = (27)^{-1/2} (L - i N)$, the QN wavefunction may be rewritten in the following form:
\beq
\tilde{u}_{ln}(r) \approx U(r)  [ \rho(r) ]^n \exp\left( i L \mathcal{R}(r) / \sqrt{27} \right)   \label{unorm-approx}
\eeq
where
\begin{align}
U(r) &=x^{-1/2} e^{-r(x-1)^2 / (4 \sqrt{27}) }  \nonumber\\
&\qquad \left(\frac{1+x}{4 - 2x} \right)^{2/\sqrt{27}}  \left(\frac{1+\sqrt{3}}{x+\sqrt{3}} \right)  ,  \label{U-def}  \\
\mathcal{R}(r) &= 4 \ln \left( \frac{1+x}{4-2x} \right) - \frac{1}{2} r (x-1)^2 ,  \label{R-def} \\
\rho(r)  &= \left( 1 - \frac{3}{r} \right) \left( \frac{1+x}{4-2x} \right)^{4/\sqrt{27}} \nonumber\\
&\qquad\left(\frac{1+\sqrt{3}}{x+\sqrt{3}} \right)^2 e^{-\frac{1}{2} r (x-1)^2 / \sqrt{27} } .  \label{rho-def}
\end{align}

\subsection{Excitation factors}
The excitation factors defined in Eq.~(\ref{Bef-def}) are somewhat more difficult to obtain, since we must compute the derivative of the ingoing coefficient $\Ain$ with respect to $\omega$, and evaluate at the QNM frequency. To leading order in $L$, it is straightforward (but somewhat tedious) to compute the excitation factors by combining the expansion method with standard WKB techniques. We present the full analysis in Appendix~\ref{appendix-Bef}, and here merely quote the key result:
\beq
\B_{ln} \approx  i^{1/2} L^{-1/2} B e^{2 i \zeta L / \sqrt{27}}  \frac{ \left( -i \kappa L \right)^n }{ n! }  \label{Bef-approx}
\eeq
where the `geometric constants' are
\begin{eqnarray}
\zeta  	&=& 3 - \sqrt{27}  + 4 \ln 2 - 6 \ln (2 + \sqrt{3}) \nonumber\\
&\approx&  -7.325311084 ,  \label{y-def}  \\
B 	&=& \frac{\sqrt{27} \, e^{\zeta / \sqrt{27}}}{(2 + \sqrt{3})\sqrt{\pi}}  \approx 0.19182703317 ,  \\
\kappa &=&  \frac{216 \, e^{2\zeta / \sqrt{27}}}{(2 + \sqrt{3})^2} \approx 0.92482482643 .  \label{kappa-def}
\end{eqnarray}

\begin{figure}
 \begin{center}
 \includegraphics[width=\columnwidth]{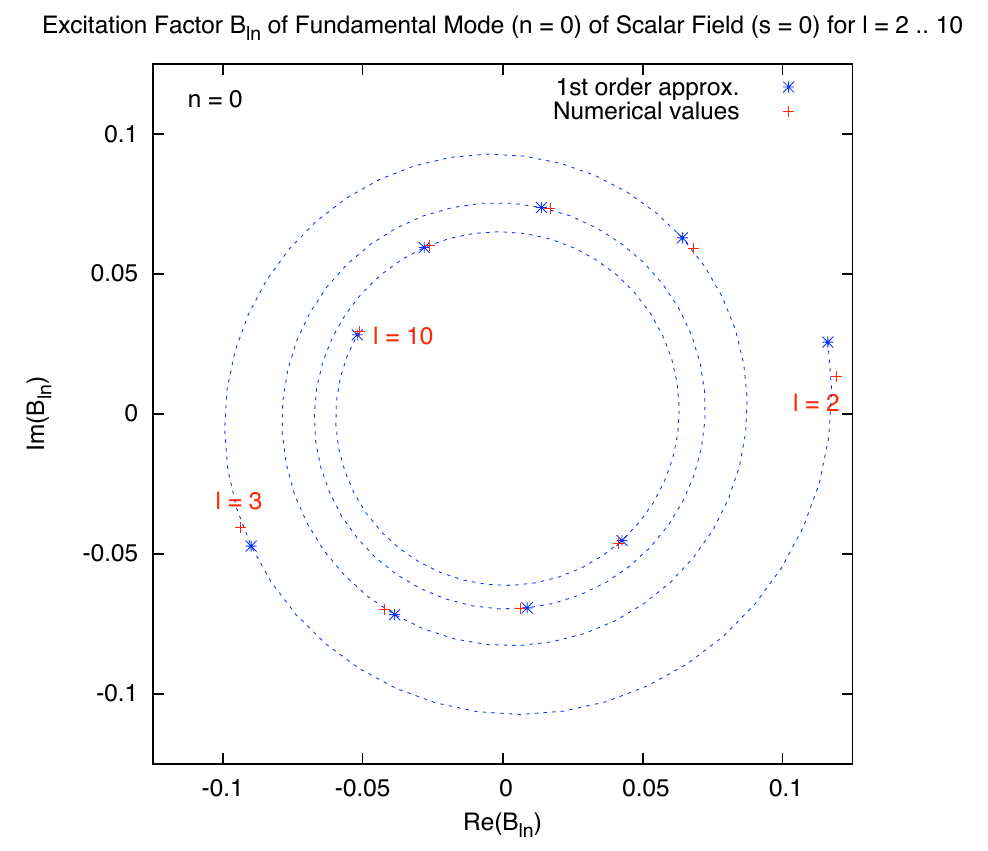}
 \caption{\emph{Excitation factors $\B_{ln}$ of fundamental ($n=0$) modes of scalar field}. The plot shows the real and imaginary parts of the excitation factors $\B_{l0}$ for modes $l = 2 \ldots 10$. The blue dotted line and markers show the asymptotic approximation [Eq.~(\ref{Bef-approx})]. The red crosses show the numerically-determined values (see e.g.~\cite{Berti-Cardoso-2006}).}
 \label{fig:ef-spiral}
 \end{center}
\end{figure}

Fig.~\ref{fig:ef-spiral} shows the `fundamental mode' ($n=0$) excitation factors for the scalar field, in the complex plane.  It compares the lowest-order large-$l$ approximation given in Eq.~(\ref{Bef-approx}) [blue] with numerically-determined excitation factors [red], for modes $l=2 \ldots 10$. The agreement is good, and improves as $l$ is increased, as expected.

\section{Structure of the Green function\label{sec:structure}}
In this section we employ the large-$l$ approximations of the previous section to investigate the singular structure of the Green function; in other words, the form of the GF close to the lightcone. Let us begin by rewriting the QNM sum (\ref{G-QNM-def}) as 
\beq
\GQNM(x,x') = \frac{2}{r r'} \text{Re} \sum_{l=0}^\infty \mathcal{G}_{l}(x,x'),  \label{Gl-def}
\eeq
where
\beq
\mathcal{G}_l(x,x') = \sum_{n=0}^\infty (2l + 1) P_l(\cos \gam) \mathcal{B}_{ln} \tilde{u}_{ln}(r) \tilde{u}_{ln}(r') e^{-i \omega_{ln} T}     \label{Gl}
\eeq
Note that the series (\ref{Gl-def}) is divergent, in the sense that $\lim_{l \rightarrow \infty} |\mathcal{G}_l| \neq 0$. In the large-$l$ limit (and for fixed points $x$, $x'$) the phase of $\mathcal{G}_l$ varies linearly with $l$, i.e. $\lim_{l\rightarrow \infty} \arg ( \mathcal{G}_{l+1} / \mathcal{G}_l ) = \text{const}$. In Sec.~\ref{sec-caustics} we show that, in general, well-defined values can be extracted from the series (\ref{Gl-def}) using an appropriate summation method \cite{Hardy}. However, the summation method does not yield a finite result if the terms $\mathcal{G}_l$ combine coherently as $l \rightarrow \infty$. In this case, the series representation of the Green function is singular.

\subsection{Singular structure and the light cone\label{sec-singularity-times}}
As argued above, the QN mode sum for the Green function is singular if the coherent phase condition is satisfied:
\beq
\lim_{l \rightarrow \infty} \arg \left(  \mathcal{G}_{l+1} / \mathcal{G}_l  \right) =  2 \pi k ,  \quad \quad k \in \mathbb{Z} .\label{coherent-phase}
\eeq
From physical considerations, we expect the Green function to be singular on the light cone (i.e.~if the spacetime points $x$ and $x'$ are connected by a null geodesic). Hence we expect to find that the phase of $\mathcal{G}_l$ is related to the null geodesics on the black hole spacetime. Below, we use the asymptotic results of Sec.~\ref{sec:asymptotics} to demonstrate the link explicitly for the Schwarzschild spacetime.

Inserting asymptotic results (\ref{unorm-approx}) and (\ref{Bef-approx}) into (\ref{Gl-def}) we obtain that, at leading order in $L = l+1/2$, 
\begin{align}
\label{Glapprox}
&\lim_{l \rightarrow \infty} \mathcal{G}_l = 2 B \, U(r) U(r') e^{-T/(2 \sqrt{27})} (i L)^{1/2}\nonumber\\
&\quad\times e^{ i L [ (2\zeta + \mathcal{R}(r) + \mathcal{R}(r') - T)/\sqrt{27} ] }P_{L-\frac12} (\cos \gamma) \nonumber \\ 
&\qquad \times   \sum_{n=0}^\infty \frac{\left[ \left(- i \kappa \rho(r) \rho(r') L e^{-T/\sqrt{27}} \right)^n + \mathcal{O}(L^{n-1}) \right]}{n!}   .
\end{align}
If we are prepared to neglect all subdominant terms in $L$ then the sum over overtones $n$ may be performed, 
\begin{align}
\label{n-sum}
\sum_{n=0}^\infty &\frac{\left(- i \kappa \rho(r) \rho(r') L e^{-T/\sqrt{27}} \right)^n}{n!} = \nonumber\\
& \qquad \exp \left( - i \kappa \rho(r) \rho(r') L e^{-T/\sqrt{27}}  \right) .
\end{align}
Let us now assume that $\gamma \neq 0, \pi$, and employ the approximation 
\begin{align}
\lim_{l \gam \gg 1} & P_l(\cos \gam) \sim\nonumber \\
& \left( \frac{1}{2 \pi L \sin \gam} \right)^{1/2} \left( e^{-i\pi/4} e^{ i L \gam} + e^{i \pi /4} e^{-i L \gam}  \right)    \label{Leg-approx}
\end{align}
With results (\ref{Glapprox}), (\ref{n-sum}) and (\ref{Leg-approx}), the coherent phase condition (\ref{coherent-phase}) is equivalent to the implicit equation
\begin{align}
T &= \sqrt{27} \left( 2 \pi k \pm \gam \right) + 2 \zeta + \mathcal{R}(r) + \mathcal{R}(r') \nonumber \\
&\qquad\qquad  - \sqrt{27} \kappa \rho(r) \rho(r') e^{-T / \sqrt{27}} ,  \label{T-sing2}
\end{align}
where $k$ is an integer. Quantities $T, \zeta, \kappa, \mathcal{R}(r), \rho(r)$ were defined in Eqs.~(\ref{reflection-time}), (\ref{y-def}), (\ref{kappa-def}), (\ref{R-def}), (\ref{rho-def}), respectively. The quantity $2 \pi k \pm \gam$ (where $0 \le \gam < \pi$ is the angle between the spacetime points) may be interpreted as the angle traversed by a null geodesic in an orbit around the black hole (see Fig.~\ref{fig:lightcone3d}).  

We expect the solutions of Eq.~(\ref{T-sing2}), i.e.~$T = T_{k\pm}^{\text{(QNM)}}(r, r', \gam)$, to be approximations to the times at which the Green function is singular; and on physical grounds we expect the Green function to be singular if (and only if) $x$ and $x'$ are connected by a null geodesic. Hence we may test our argument by comparing $T_{k\pm}^{\text{(QNM)}}$ against the `null geodesic times', $T_{k\pm}^{(\text{geo})}$, found by integrating the Schwarzschild orbital equations.  

Table~\ref{tbl-sing-times} compares the `QNM times' with `geodesic times', for the special case of $\gamma = 0$ and $r = r'$. In the first column is the time $T_1^{(\text{geo})} = \Delta t - 2 r_\ast$ where $\Delta t$ is the coordinate time it takes for a null geodesic to start at a point, orbit once around the black hole, and return to the same point (i.e. $r=r', \gam = 0$). In the second column is the time $T_1^{\text{(QNM)}}$ found by solving Eq.~(\ref{T-sing2}), using a simple iterative method. We find excellent, though not precise, agreement. The `QNM time' approaches the `geodesic time' as the number of orbits goes to infinity (i.e.~$k \rightarrow \infty$). 

\begin{table}[h]
\begin{tabular}{l | r r r}
& $T_1^{\text{(geo)}}$ & $T_1^{\text{(QNM)}}$ & $T_1^{\text{(geo)}} - T_1^{\text{(QNM)}} $ \\
\hline
$r=4$ &  \quad  \quad $25.6449315$& \quad \quad $25.6449191$ & \quad \quad $0.00001246$ \\
$r=10$ & $20.7019188$ & $20.7017749$ & $0.00014394$ \\
$r=100$ & $18.1280305$ & $18.1285177$ & $-0.00048718$ 
\end{tabular}
\caption{\emph{Singularity Times}.   Here $T= \Delta t - 2\rstar$ where $\Delta t$ is the time taken for a null geodesic starting at radius $r$ to orbit the black hole once and return to the starting radius. $T^{\text{(QNM)}}$ is found by solving the implicit equation (\ref{T-sing2}), and $T^{\text{(geo)}}$ is found by solving the geodesic equations numerically. }
\label{tbl-sing-times}
\end{table}



\begin{figure}
 \begin{center}
 \includegraphics[width=\columnwidth]{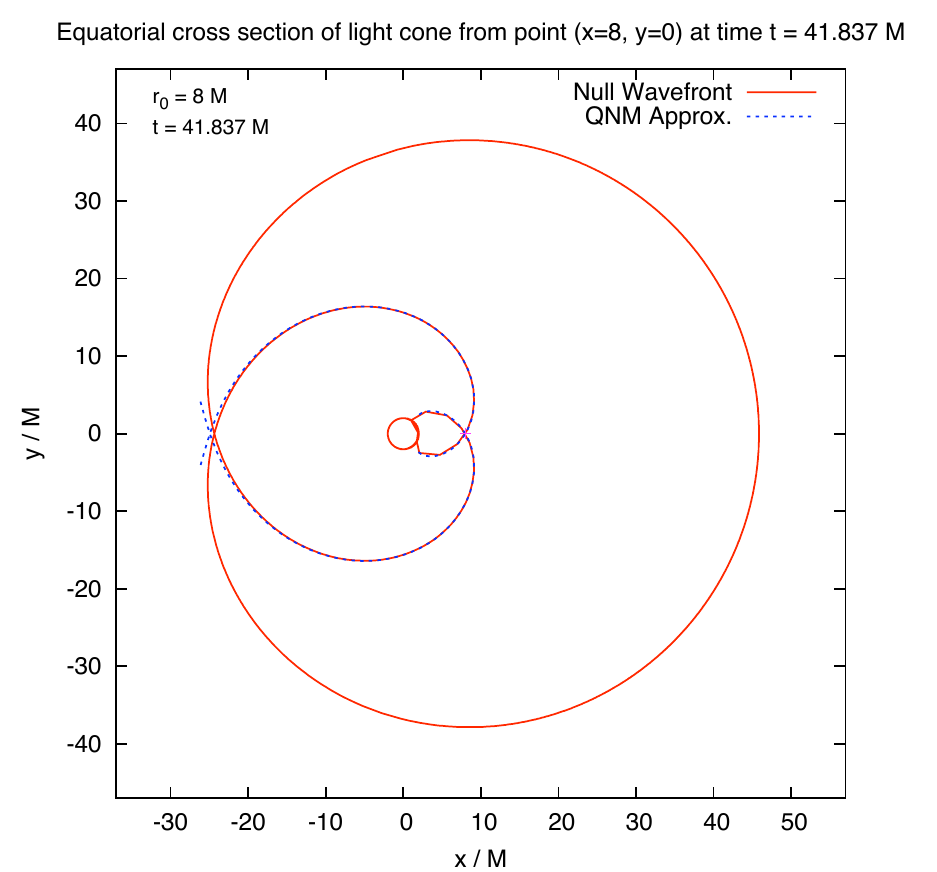}
 \caption{\emph{Wavefront cross section}. This plot shows a `snapshot' in the equatorial plane of a wavefront [red solid line] emanating from $x'=8M, y'=0$, $t'=0$, taken at coordinate time $t = t_1 \approx 41.8M$. It corresponds to a cross section of the light cone shown in Fig.~\ref{fig:lightcone3d}. Note that the wavefront intersects itself at two antipodal points (caustics). The blue dotted line shows the position of the wavefront predicted using the large-$l$ asymptotics of the QNMs [Eq.~(\ref{T-sing2})]. See also Fig.~\ref{fig:lightcone-gf}.}
\label{fig:snapshot}
 \end{center}
\end{figure}

Fig.~\ref{fig:snapshot} shows a cross section [in the plane $\theta=\pi/2$] of the wavefront at $t = t_1$ associated with an initial event at $x' = 8M$, $y' = t' =0$ (see also Fig.~\ref{fig:snapshot}). Here we have chosen $t_1 \approx 41.8M$, which is the time is takes for the wavefront to orbit once and return to $x=8M$. This is sufficient time for two caustics to form. The dotted blue line shows the prediction of the position of the wavefront found from inserting the QNM asymptotics of Sec.~\ref{sec:asymptotics} into the coherent phase condition Eq.~(\ref{coherent-phase}) [more explicitly, for a given $k, \pm, \gam$ we solved (\ref{T-sing2}) numerically and rearranged (\ref{reflection-time}) to find $r_\ast$ and hence $r$].  The plot shows that the wavefront prediction is most accurate for large $T$, in other words, close to the unstable orbit at $r=3M$. 

It is natural to presume that the (exact) QN mode sum (\ref{Gret-mode-sum}) is in fact singular \emph{precisely} on the light cone (and not just close to it). That is, we presume that the difference between the QNM time determined from Eq.~(\ref{T-sing2}) and the geodesic time must arise because we have neglected important sub-dominant terms in the expansions in $L$. In particular, performing the sum over $n$ in Eq.~(\ref{n-sum}) is a non-rigorous step that should be viewed with some suspicion. It may be  possible to improve Eq.~(\ref{T-sing2}) somewhat by extending the approximations (\ref{unorm-approx}) and (\ref{Bef-approx}) to higher order in $L$.

\subsection{Caustics and the four-fold singular structure\label{sec-caustics}}
In the previous subsection we demonstrated the link between the light cone and the large-$l$ asymptotics of the terms in the QN mode sum (\ref{G-QNM-def}). Here we make further use of the large-$l$ asymptotics to investigate the form of the scalar-field Green function close to the light cone. We show that the singular part of the Green function undergoes a transition each time a caustic is encountered (where we use the term caustic to refer to a self-intersection of the light cone).

Let us start by inserting the large-$l$ asymptotics of Sec.~\ref{sec:asymptotics} into the QNM decomposition of the Green function (\ref{G-QNM-def}). We obtain the expression 
\begin{align}
\Gsing(x,x') &= \mathcal{X}(r,r',T) \, \nonumber\\
&\text{Re} \sum_{l=0}^\infty (i L)^{1/2} P_l (\cos \gam) e^{-i L \Phi(r,r',\gam,T)}
\label{G-sing-def}
\end{align}
where
\begin{align}
\mathcal{X}(r,r',T) &= \frac{4 B \, U(r) U(r') e^{-T/[2\sqrt{27}]}}{r r'} 
\end{align}
and
\begin{align}
\Phi(r,r',\gam,T) &= \frac{T - \mathcal{R}(r) - \mathcal{R}(r') - 2\zeta} {\sqrt{27}} \nonumber\\
  &\qquad\qquad+ \kappa \rho(r) \rho(r') e^{-T/\sqrt{27}} .
\end{align}
Here $\Gsing$ is a quantity which should have the same singular structure as $\GQNM$ and (we assert) $\Gret$.

We now follow the steps taken in Sec.~VD of \cite{CDOW1}, where a similar analysis was performed for the Nariai spacetime \cite{Nariai}. Related methods are used in, e.g., seismology \cite{Aki-Richards}. First, we make use of the Poisson sum formula to convert the infinite sum (\ref{G-sing-def}) into an integral, viz.,
\begin{widetext}
\begin{eqnarray}
G_{ret}^{(\text{sing})} &=& \mathcal{X} \sum_{s=-\infty}^{+\infty} (-1)^s \text{Re} \int_{L=0}^\infty dL \, e^{2\pi i s L} (iL)^{1/2} e^{-iL\Phi} P_{L-1/2}(\cos \gam) = \mathcal{X}  \sum_{m=0}^{\infty}  \mathcal{I}_m
\label{Gret-sing}
\end{eqnarray}
where
\beq
\mathcal{I}_m = \text{Re} \int_0^\infty dL (iL)^{1/2} e^{-iL\Phi} R_m   \label{I_N-def}
\eeq
and
\beq
R_m = \left\{  \begin{array}{ll} (-1)^{m/2} \left( \mathcal{Q}^{(-)}_{L-1/2}(\cos \gam) e^{i m \pi L} +  \mathcal{Q}^{(+)}_{L-1/2}(\cos \gam) e^{-i m \pi L} \right), & m \; \text{even}, \\
\\
(-1)^{(m+1)/2} \left( \mathcal{Q}^{(+)}_{L-1/2}(\cos \gam) e^{i (m+1) \pi L} +  \mathcal{Q}^{(-)}_{L-1/2}(\cos \gam) e^{-i (m+1) \pi L} \right), & m \; \text{odd},
 \end{array} \right.
\eeq
with
\beq
\mathcal{Q}^{(\pm)}_\mu(\cos \gam) \equiv \frac{1}{2} \left( P_{\mu}(\cos \gam) \pm \frac{2i}{\pi} Q_\mu(\cos \gam)  \right) ,
\eeq
\end{widetext}
where $Q_\mu(\cdot)$ is the Legendre function of the second kind. The integrals (\ref{I_N-def}) can be approximated if we make use of asymptotic approximations for the $\mathcal{Q}^{\pm}_{L-1/2}$ functions which are valid in the large-$l$ limit. We will make use of the following simple `exponential' approximation:
\beq
\mathcal{Q}_{L-1/2}^{(\pm)} (\cos \gam) \approx \left( \frac{1}{2 \pi L \sin \gam} \right)^{1/2} e^{\pm i \pi / 4} e^{\mp i L \gam} .  \label{Q-approx-1}
\eeq
This approximation is valid in the regime $L \gam \gg 1$ (and $L |\pi - \gam| \gg 1$); in other words, it is inappropriate for use if the spacetime points are separated by a small angle $\gam \sim 0$, for example, close to a caustic.  

Applying approximation (\ref{Q-approx-1}) we obtain
\begin{widetext}
\beq
\mathcal{I}_m \approx \left( \frac{1}{2 \pi \sin \gam} \right)^{1/2}  \text{Re} \int_{L=0}^\infty  dL
\left\{ \begin{array}{ll}
  (-1)^{m/2} \left[ e^{i L (m\pi + \gam - \Phi)} + i e^{-iL(m\pi + \gam + \Phi)} \right],  & m \; \text{even}, \\
  (-1)^{(m+1)/2} \left[ i e^{i L ( [m+1] \pi - \gam - \Phi)} + e^{-iL([m+1]\pi - \gam + \Phi)} \right],  & m \; \text{odd},
\end{array} \right.  \label{eq-Im}
\eeq
\end{widetext}
for $L \gg 1$. The integral will be singular if the phase factor in either term in (\ref{eq-Im}) is zero; unsurprisingly, this is equivalent to the coherent phase condition we investigated in Sec.~\ref{sec-singularity-times}. The QN mode sum (\ref{Gret-mode-sum}) is only expected to be convergent for $T > 0$ (hence $\Phi > 0$) and so for simplicity let us neglect the latter terms in (\ref{eq-Im}). Now we may regularize the integral by introducing a small positive imaginary part in the exponent [introducing a small $\epsilon$ into the exponent of (\ref{eq-Im}) is equivalent to imposing a smooth high-$l$ cutoff upon the integral], and use the identity
\beq
\lim_{\eps \rightarrow 0^+} \int_0^\infty  e^{i L (q + i \epsilon)} dL = \frac{i}{q} + \pi \delta(q) ,
\eeq
to obtain
\beq
\mathcal{I}_m \approx \left( \frac{\pi}{2 \sin \gam} \right)^{\!\! 1/2} 
\left\{ \begin{array}{ll}
 (-1)^{m/2} \, \delta( m \pi + \gam - \Phi ), & m \; \text{even},  \\
\displaystyle  \frac{\phantom{{^{(m+1)/2}}}(-1)^{(m+1)/2}}{ \pi [\Phi - (m+1)\pi + \gam ] }, & m \; \text{odd}.  \\ 
\end{array} \right.  \label{Im}
\eeq
From the above expression, we see that the sum is singular when $\Phi = 2 \pi k \pm \gam$ (where $k$ is an integer), as anticipated in Sec.~\ref{sec-singularity-times} [see Eq.~(\ref{T-sing2})]. However, the form of the contribution $\mathcal{I}_m$ to the Green function depends crucially on whether $m$ is even or odd. The `shape' of the singularity alternates between a delta distribution and a singularity with antisymmetric `wings'. The $m$th wave may be associated with the $m$th orbiting null geodesic. Now, $m$ has a clear geometrical interpretation: it is the number of caustics (antipodal points) through which the corresponding geodesic has passed. Equation (\ref{Im}) implies that the Green function changes form each time the wave front intersects itself (i.e.~passes through a caustic). 

The simple argument above gives us some insight into the `four-fold' singular structure of the Green function \cite{Ori1}. However, the approximation (\ref{Q-approx-1}) leading to (\ref{Im}) breaks down when $\gamma$ is close to $0$ or $\pi$. For example, the approximation is not valid close to a caustic. To improve our analysis, we may employ the uniform asymptotics established by Olver \cite{Olver}, 
\beq
\mathcal{Q}_{L-1/2}^{(\pm)} (\cos \gam) \approx  \frac{1}{2} \left( \frac{\gam}{\sin \gam} \right)^{1/2} H_0^{\mp}(\gamma L) ,
\label{Q-approx-2}
\eeq
where $H^{\pm}_0(\cdot) = J_0(\cdot) \pm i Y_0(\cdot) $ are Hankel functions of the first ($+$) and second ($-$) kinds. The asymptotic approximation (\ref{Q-approx-2}) is valid for $l \gg 1$ for all angles $\gam < \pi$. Using (\ref{Q-approx-2}) in (\ref{I_N-def}) leads to 
\begin{widetext}
\beq
\mathcal{I}_m \approx \left( \frac{\gam}{4 \sin \gam} \right)^{\!\! 1/2} 
\! \text{Re} \int_0^\infty \!\! dL (iL)^{1/2} 
\left\{ \begin{array}{ll}
 (-1)^{m/2} \left[ H^{+}_0(\gam L) e^{iL(m\pi - \Phi)} + H^{-}_0(\gam L) e^{-iL(m\pi + \Phi)} \right]  , & m \; \text{even},  \\
\\
 (-1)^{(m+1)/2} \left[ H^{-}_0(\gam L) e^{iL([m+1]\pi - \Phi)} + H^{+}_0(\gam L) e^{-iL([m+1]\pi + \Phi)} \right] , & m \; \text{odd}.  \\ 
\end{array} \right.  \label{Im2}
\eeq
\end{widetext}
Such integrals can be found with the aid of results in standard tables books, for example, Gradshteyn \& Ryzhik~\cite{Gradshteyn-Ryzhik}. Once again, for simplicity let us neglect the latter terms (which are singular for $T < 0$). Consider now the even-$m$ case; the first term is singular when $\Phi = m\pi + \gamma$. If $\Phi < m\pi + \gamma$, we may integrate along the positive imaginary axis by making the substitution $L = i \nu$, and using the identity $H_0^+(i\gamma\nu) = 2 K_0(\gamma \nu) / (i \pi)$. The integral is purely-imaginary, and hence upon taking the real part we get zero. Conversely, if $\Phi > m\pi + \gamma$, then we may integrate along the negative imaginary axis by making the substitution $L = -i\nu$ and using the identities $H_0^+(-i\gam \nu) = 2I_0(\gam \nu) - H_0^-(-i\gam \nu) = 2I_0(\gam \nu) + 2K_0(\gam \nu) / (i \pi)$. Taking the real part picks out only the latter term, and the integral may be found with the aid of Eq.~6.621(3) in \cite{Gradshteyn-Ryzhik}. Proceeding in this way, we obtain
\begin{widetext}
\begin{eqnarray}
\mathcal{I}_{m} = 
\frac{1}{2}\left( \frac{\gam}{\sin \gam} \right)^{1/2}
\begin{cases} 
(-1)^{m/2} \left\{ \begin{array}{ll}
 \left(\frac{2 \pi}{\gamma} \right)^{1/2} \delta\left(  \Phi - [m \pi + \gam] \right) , & \Phi \le m \pi + \gam, \\
{\displaystyle  -\frac{\pi^{1/2}}{2[\Phi - m \pi + \gam ]^{3/2}} } {}_2F_1\left(\frac{3}{2}, \frac{1}{2}; 2; \frac{\Phi - m\pi - \gam}{\Phi - m\pi + \gam} \right), & \Phi > m \pi + \gam,
\end{array} \right.
 &  m \text{ even},  \\
\\
 (-1)^{{(m+1)}/{2}} \left\{ \begin{array}{ll}
 \displaystyle\frac{-2E \left(2 \gam / [(m+1)\pi + \gam - \Phi] \right)}{\pi^{1/2} [(m+1)\pi - \gam - \Phi][(m+1)\pi + \gam - \Phi]^{1/2}},
& \Phi < (m+1)\pi - \gam , \\
{\displaystyle \frac{\pi^{1/2}}{2[\Phi - (m+1)\pi + \gam]^{3/2}}} {}_2F_1 \left( \frac{3}{2}, \frac{1}{2} ; 2 ; \frac{\Phi - (m+1)\pi - \gam}{\Phi - (m+1)\pi + \gam} \right) ,
& \Phi > (m+1)\pi - \gam ,
\end{array} \right.
 & m \text{ odd}.
\end{cases}
\label{GF-asymp-result-2}
\end{eqnarray}
\end{widetext}
Here ${}_2F_1$ is the Gauss hypergeometric function, and $E$ is the complete elliptic integral of the second kind. Note that the delta function deduced in Eq.~(\ref{Im}) has been included in the top line of Eq.~(\ref{GF-asymp-result-2}). 
Eq.~(\ref{GF-asymp-result-2}) is valid in the range $0 \le \gam \lesssim \pi/2$; but extending the result to $\pi/2 \lesssim \gam \le \pi$ is straightforward. 

\subsection{Numerical visualisations}
\begin{figure}
 \begin{center}
 \includegraphics[width=\columnwidth]{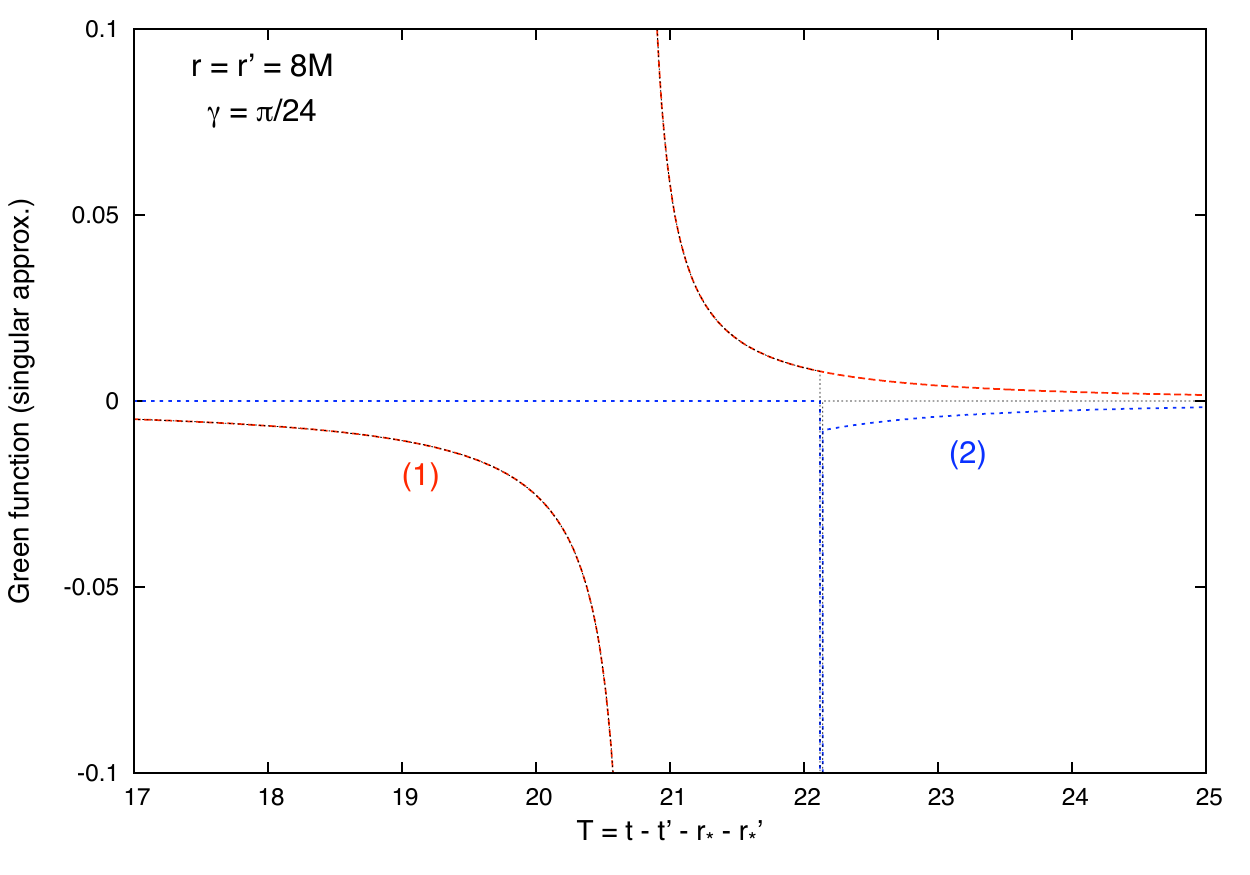}
 \caption{\emph{Asymptotic approximation to the singular structure of the Green function.} This plot illustrates the GF as a function of $T = t - t' - \rstar - \rstar'$, for fixed spatial points at $r = r' = 8M$ separated by a small angle, $\gamma = \pi / 24$. The black dashed line was computed using approximation to the Green function given by Eq.~(\ref{Gret-sing}) and (\ref{GF-asymp-result-2}). The red curve (1) shows the contribution from $\mathcal{I}_1$, and the blue curve (2) shows the contribution from $\mathcal{I}_2$. The `spike' at $T \approx 22.13$ indicates a (-ve) delta function. In the regime $T > 22.13$, the terms $\mathcal{I}_1$ and $\mathcal{I}_2$ cancel precisely. 
 }
 \label{fig:I1I2}
 \end{center}
\end{figure}

Figure \ref{fig:I1I2} shows the asymptotic approximation (\ref{GF-asymp-result-2}), for fixed points separated by angle $\gamma = \pi/24$ at radius $r = r' = 8M$ as a function of reflection time $T = t - t' - r_\ast - r_\ast'$. The contributions from $\mathcal{I}_1$ and $\mathcal{I}_2$ are shown separately. The GF is singular at $T = T_1, T_2, \ldots$, where $T_1$ and $T_2$ correspond to the time taken for a null geodesic to pass through an angle $2\pi - \gam$ and $2\pi + \gam$, respectively. The former geodesic (1) passes through one caustic, and the latter (2) through two. Near $T_1$ the GF is approximately antisymmetric. At $T_2$ the GF has delta-function support. Beyond $T > T_2$ the contributions from $\mathcal{I}_1$ and $\mathcal{I}_2$ cancel precisely. 

\begin{figure}
 \begin{center}
  \includegraphics[width=\columnwidth]{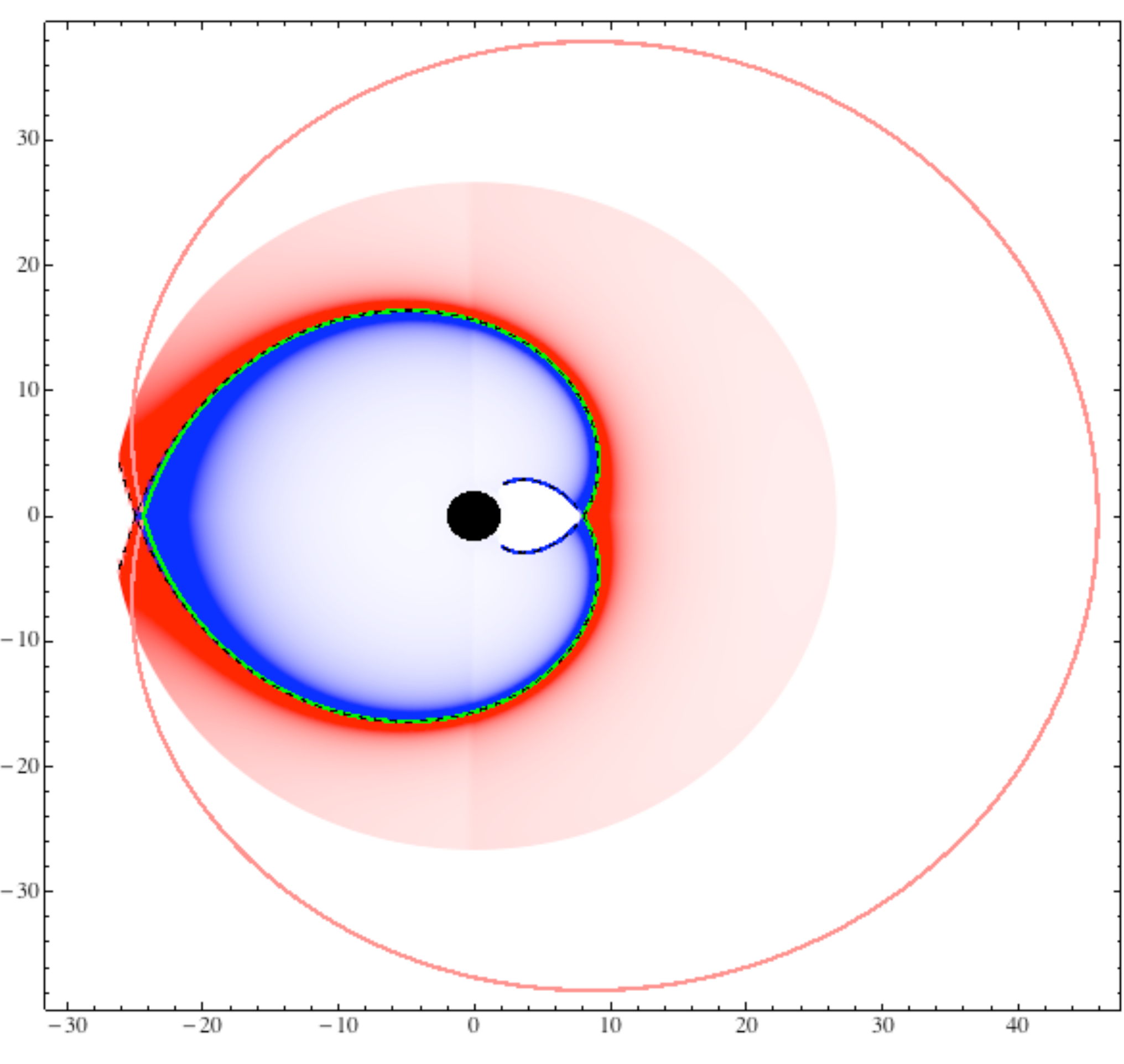}
 \caption{\emph{Illustration of the singular structure of the Green function.} This plot shows a `snapshot' of the scalar-field Green function (GF) in the $xy$ plane at $t = t_1$ associated with the initial point at $x' = 8M, y' = 0, t' =0$. Note that $t_1 \approx 41.837M$ is the coordinate time it takes for part of the initial wavefront to orbit the BH once and return to $x=x'$, $y=0$. On the outermost wavefront [pink line], which lies outside the first caustic point at $x \approx -24.36M, y = 0$, the GF is proportional to a positive delta function [Eq.~(\ref{eq-Im})]. On the innermost wavefront [blue line], inside the second caustic point (at $x = 8M, y = 0$), the GF is proportional to a negative delta function. Across the intermediate wavefront [green line] which joins the two caustics, the GF has antisymmetric `wings' approximated by $1/\sigma$ (where $\sigma$ is the Synge worldfunction). The dotted black line shows the position of the wavefront estimated from the lowest-order asymptotics of the QNM sum (see also Fig.~\ref{fig:snapshot}). The red (positive) and blue (negative) shadings indicate the value of the GF (computed via the asymptotic approximations [Eq.~(\ref{GF-asymp-result-2})]). The coloured disk indicates the area of spacetime in which the sum over QNM overtones [Eq.~(\ref{Gl})] is convergent.}
 \label{fig:lightcone-gf}
 \end{center}
\end{figure}

The structure of the GF in the $xy$ plane (i.e.~$\theta = \pi/2$) is illustrated in Fig.~\ref{fig:lightcone-gf}. It shows a snapshot at $t = t_1$ of the wavefront [solid lines] and GF [shading] associated with an event at $x' = 8M$, $y' = t' =0$ (see also Fig.~\ref{fig:snapshot}). The first and second caustics are visible (at $x \approx -24.36M$ and at $x = 8M$). Observe that different parts of the wavefront have passed through zero (pink line, outer), one (green) and two (purple) caustics. The plot illustrates how the GF changes depending on the number of caustics crossed. On the outer wavefront (0 caustics, pink), the GF has delta-function support. Around the intermediate wavefront (1 caustic, green), the GF is antisymmetric (positive on the exterior, negative on the interior). On the inner wavefront (2 caustics, purple), the GF again has delta-function support, but with the opposite (-) sign.

\section{Discussion and Conclusion\label{sec:conclusion}}

In this paper we have used the expansion method of \cite{Dolan-Ottewill} to investigate the large-$l$ asymptotics of the Schwarzschild QN modes and their excitation factors. We used the asymptotic results to investigate the QNM contribution to the scalar-field retarded Green function. This led to insight into the singular structure of the GF near the lightcone. We showed that the form of the GF changes every time a caustic is encountered, and that it undergoes a four-fold repeating pattern [see Eq.~(\ref{eq-Im}), (\ref{Im}), (\ref{GF-asymp-result-2}) and Fig.~\ref{fig:lightcone-gf}].

A four-fold singular structure for a 4D spherically-symmetric spacetime (like the Schwarzschild black hole) has been anticipated (e.g.~\cite{Ori1, CDOW1}); however, we believe this is the first time it has been demonstrated for a black hole spacetime using a QN mode representation. As was discussed in Sec.~VE of Ref.~\cite{CDOW1}, an alternative way to make sense of the four-fold pattern is with a Hadamard ansatz \cite{Hadamard} for the `direct' part of the GF, i.e.~
\begin{eqnarray}
\Gret^{\text{dir.}}(x,x') &\sim& \lim_{\epsilon \rightarrow 0^+}  \text{Re} \left[ i \frac{ \Delta^{1/2}}{ \pi \sigma + i \epsilon} \right] \nonumber \\
 &\sim& \text{Re} \left[ \Delta^{1/2} \left( \delta(\sigma) + \frac{i}{\pi \sigma} \right) \right] . \label{hadamard}
\end{eqnarray}
Here $\sigma = \sigma(x,x')$ is the Synge world function \cite{Synge} (half the squared interval along the geodesic joining $x$ to $x'$), and $\Delta = \Delta(x,x')$ is the van Vleck determinant, a measure of the degree of focussing of neighbouring geodesics \cite{OW2}. The determinant is singular at the caustic. In passing through a caustic on the Schwarzschild spacetime, $\Delta$ changes sign. If we presume that this leads to an additional factor of $-i$ in the square bracket of (\ref{hadamard}) each time a caustic is traversed, we recover once more the four-fold pattern $\delta(\sigma)$, $1/(\pi \sigma)$, $-\delta(\sigma)$, $-1/(\pi \sigma)$, $\delta(\sigma)$, etc.

The effect of caustics upon wave propagation has been studied with a variety of methods (e.g. WKB methods, path integrals, functional integration, etc.) in a range of disciplines including optics \cite{Stavroudis}, acoustics \cite{Kravtsov}, seismology \cite{Aki-Richards}, symplectic geometry \cite{Arnold}, and quantum mechanics \cite{Berry-Mount, Schulman}. The formation of caustics in black hole spacetimes has also been studied extensively, in the context of strong gravitational lensing \cite{Perlick, Bozza-2008, Bozza-Mancini-2009}. Some further investigation of the effect of caustics on the details of wave propagation (i.e.~on the Green function) in curved spacetimes now seems to be warranted. In particular, analysis of the singular structure of the Green function near a rotating (Kerr) black hole, an axisymmetric spacetime, would be of great interest \cite{Tamburini}. For this challenge, the utility of the QN expansion method appears rather limited \cite{Dolan-2010}. Instead, let us focus (for now) on what more can be done in Schwarzschild with our methods. 

In Paper I \cite{Dolan-Ottewill} we introduced the QNM expansion method and applied it to investigate the QN frequency spectrum. In this work -- Paper II, let's say -- we have explored the geometric interpretation, using first-order (in $L$) expansions to bring \emph{qualitative} understanding of the singular structure of the GF. It seems there is scope for a future work (i.e.~Paper III) in which higher-order (analytic) expansions are combined with numerical methods \cite{Leaver-1985, Berti-Cardoso-2006} in the small-$l$ regime, to achieve \emph{quantitative} calculations of QNM sums [such as Eq.~(\ref{G-QNM-def})]. 

Let us suggest two areas where this work may find further applications. The first is in the calculation of the gravitational self-force \cite{Poisson, CDOW1}, which is required to (e.g.) model extreme mass-ratio black hole inspirals \cite{Barack}, test the cosmic censorship hypothesis \cite{Barausse}, and calibrate effective one-body theories \cite{Damour}. With complete (i.e.~global) knowledge of the GF, it is straightforward (in principle) to compute the self-force; yet the former is not trivial to obtain. In the `method of matched expansions' \cite{Anderson-Wiseman} one seeks to match a `quasilocal' expansion of the GF \cite{OW1}, valid inside a normal neighbourhood (approximately, up to the first caustic), to a `distant-past' expression. The QNM expansion makes up part of the latter; however, it must also be augmented by an accurate calculation of the branch cut integral. This remains to be achieved.

A second possible application is to the theory of Complex Angular Momentum (CAM) on black hole spacetimes \cite{DF, DFR, DEFF, DFR2}. It is well-established (see e.g.~\cite{Dolan-2007}) that the black hole photon orbit creates regular interference features in scattering cross sections, that is, `spiral scattering' oscillations at intermediate angles and a `glory' in the backward direction. In \cite{Andersson94} it was shown that such oscillations are intimately linked to the Regge poles of CAM theory. Regge poles are the complex-$l$ versions of QNMs. Very recently, D\'ecanini and Folacci \cite{DecFol} have extended the expansion method to obtain high-frequency asymptotics for the Regge pole residues (which, in CAM theory, play an equivalent role to the QNM excitation factors $\B_{ln}$). We anticipate that their work will lead to improved analytic estimates for spiral/glory scattering. 

In conclusion, we hope that our study of the large-$l$ asymptotics of QN modes has shed some light on the global structure of the Green function on a black hole spacetime, and we anticipate further illumination from work to come.

\begin{acknowledgments}
SRD acknowledges financial support from the Engineering and Physical
Sciences Research Council (EPSRC) under grant no. EP/G049092/1; ACO acknowledges financial support from 
Science Foundation Ireland (SFI) under grant no. 10/RFP/PHY2847. We would like to thank Marc Casals, Barry Wardell, Leor Barack, Kirill Ignatiev, Emanuele Berti and Vitor Cardoso for interesting discussions which influenced this work. In particular, we are grateful to Barry Wardell for his assistance in producing Fig.~\ref{fig:lightcone-gf}. 
\end{acknowledgments}

\appendix

\section{QNM excitation factors $\Bef$ via WKB analysis\label{appendix-Bef}}

In this section we derive a leading-order (in $L$) approximation to the excitation factors, defined in Eq.~(\ref{Bef-def}). The key ingredient in the calculation is the derivative $\partial \Ain /  \partial \omega$, taken at the QNM frequency. To find this, we must perturb the frequency slightly away from its QNM value, $\omega_{ln} = \bar{\omega}_{ln} + \epsilon$, (where $\epsilon$ is small) to determine the first-order change in $\Ain$. Note that we use overbar notation $\bar{\omega}$ to denote the precise QNM value. In other words we perturb only $\varpi_0$, the $\mathcal{O}(L^0)$ term in expansion (\ref{omega-series}), so that
\beq
\varpi_0 = -i N / \sqrt{27}  + \epsilon . \label{om1-perturb}
\eeq
This results in a breakdown of continuity at $r=r_c=3$ in expansion (\ref{v-expansion}). However, we may find regular solutions in a `interior' region close to $r=3$, and match these onto `exterior' solutions (in regimes $r > 3 + \epsilon$ and $r < 3 - \epsilon$), to obtain $\Ain = \mathcal{O}(\epsilon)$ to first order. In short, we combine the new expansion method with a standard WKB approach to derive the leading-order result for the excitation factors.

\subsection{Interior solution}
To find an `interior solution' valid near $r=3$, let us begin with the radial equation (\ref{rad-eq-1}) and make the ansatz
$
u_{l\omega}(r) = f^{-1/2} \psi
$
so the radial equation becomes
\beq
\frac{d^2 \psi}{dr^2} + U(r) \psi = 0 , 
\eeq
where
\beq
U(r) = f^{-2}  \left[ \omega^2 - f \left( \frac{L^2 - 1/4}{r^2} + \frac{M}{r^4} \right)  \right] .   \label{alt-de}
\eeq
Now substitute in the expansion (\ref{omega-series}) for $\omega$, and make the change of variables from $r$ to $z$ via
\beq
r = 3 + \sqrt{3} L^{-1/2} z,   \label{z-def}
\eeq
so that
\beq
U(r) = (2 \sqrt{3}\,  \varpi_0 + z^2/3 ) L + \mathcal{O} \left( L^{1/2} \right)  .
\eeq
Then, at leading order ($L^1$) the differential equation (\ref{alt-de}) implies
\beq
\frac{d^2 \psi}{d z^2} + \left( 2 \sqrt{27} \, \varpi_0 + z^2 \right) \psi = 0 .
\eeq
Now substitute in the perturbed frequency (\ref{om1-perturb}) to obtain
\beq
\frac{d^2 \psi}{d z^2} + \left( z^2 -  2 i N + 2\sqrt{27} \eps \right) \psi = 0 ,
\eeq
which has independent solutions
\beq
\psi_1 = D_{n + \eta} ( (-1+i) z) ,  \quad \quad \psi_2 = D_{n + \eta} ( (1-i) z) 
\eeq
where $\eta = \sqrt{27} i \epsilon$ and $D_\alpha(\cdot)$ is a parabolic cylinder function \cite{BenderOrzag}. 
We will use $\psi_1$, since it turns out to have the correct `ingoing' behaviour towards $z \rightarrow - \infty$. The large-$z$ asymptotics of the solution $\psi_1$ are straightforward to obtain. As $z \rightarrow -\infty$ we have
\begin{eqnarray}
\psi_1 &\sim& 2^{(n + \eta) / 2} e^{- i \pi (n + \eta) / 4} |z|^{n + \eta} e^{+i z^2 / 2} , 
\end{eqnarray}
and as $z \rightarrow + \infty$ we have
\begin{align}
\psi_1 &\sim  2^{(n+\eta)/2} e^{3 \pi i (n + \eta) / 4} |z|^{n+\eta} e^{+i z^2 / 2} \nonumber \\
            & - \frac{(2 \pi)^{1/2}}{\Gamma(-(n+\eta))} e^{i \pi (n + \eta)} e^{-3 i \pi (1 + n + \eta) / 4} \nonumber \\
  &\quad \times 2^{-(1 + \eta+n)/2} |z|^{-(\eta + n + 1)} e^{- i z^2 / 2} .
\end{align}
Hence to lowest order in $\eps$ these are simply
\beq
\psi_1  \sim 2^{n/2} e^{-i \pi n / 4} |z|^n e^{+i z^2 / 2} , \quad \quad \quad z \rightarrow -\infty,   \label{psi-asymp1}
\eeq
and
\begin{align}
\psi_1 &\sim 2^{n/2} e^{3 i \pi n / 4} |z|^n e^{+i z^2 / 2} \nonumber \\
            &  + \, \epsilon \, \sqrt{27} \, i \, \Gamma(n+1) (2 \pi)^{1/2} e^{-3 i \pi (n+1) / 4}\nonumber \\
&\qquad \times 2^{-(n+1)/2} |z|^{-(n+1)} e^{-i z^2 / 2} .   \label{psi-asymp2}
\end{align}

\subsection{Exterior solutions}
In this section we construct solutions valid in the exterior regions, i.e.~on either side of the interior around $r = 3$. Let us define solutions
\beq
u^{\pm}(r) = \exp \left(\! \pm  i \omega \int_3^{r} (1+6/r)^{1/2} (1-3/r) d\rstar  \!\right)  v^{\pm}(r)   \label{upm-def}
\eeq
where $u^{\pm}$ satisfy Eq.~(\ref{rad-eq-1}) and $v^{\pm}$ are solutions of
\begin{align}
 &f \frac{d^2 v^\pm}{dr^2} + \left[ \frac{2}{r^2}  \pm 2i\omega \left( 1+ \frac{6}{r} \right)^{1/2} \left(1 - \frac{3}{r} \right) \right] \frac{dv^{\pm}}{dr}  \nonumber \\
&\> + 
 \left[ \frac{27 \omega^2 - \lam^2}{r^2} \pm \frac{27i \omega}{r^3} \left(1 + \frac{6}{r} \right)^{\!\!-1/2}  \!+ \frac{1}{4r^2} - \frac{2\beta}{r^3} \right] v^\pm = 0 \label{rad-eq-3}
\end{align}
Note that, around the critical orbit $r=3$, we have
\begin{align}
&\exp \left( \pm  i \omega \int_3^{\rstar}  (1+6/r)^{1/2} (1-3/r) d\rstar \right) \approx\nonumber \\
  &\hspace{1.5cm}\exp( \pm i z^2 / 2  + \mathcal{O}(L^{-1})  ) .
  \end{align}
As in Eq.~(\ref{v-expansion}), we make the ansatz 
\beq
v^{\pm} = \left[ (1-3/r)^n + \mathcal{O}(L^{-1}) \right] \exp( S_{0n}^{\pm}(r) + L^{-1} S_{1n}^{\pm} + \dots).
\eeq
After plugging in to (\ref{rad-eq-3}) and matching like powers of $L$ we obtain at order $L^1$ the following:
\begin{align}
 \pm 2i &\left(1+\frac{6}{r}\right)^{1/2} \left(1 - \frac{3}{r} \right) S_{0n}^{\pm \, \prime} \pm \frac{6 i n}{r^2} \left( 1 + \frac{6}{r} \right)^{1/2} 
\nonumber\\
& \qquad \pm \frac{27 i}{r^3} \left(1+\frac{6}{r} \right)^{-1/2} + \frac{54 \omega_1}{r^2} = 0 .
\end{align}
At the QNM frequency $\varpi_0 = \bar{\varpi}_0 = -i \, N / \sqrt{27}$ the solution is $S_{0n}^+(r) = \bar{S}_{0n}^+(r)$, where
\begin{align}
&
\bar{S}_{0n}^+(r) =  \frac{1}{4} \ln \left( \frac{4r}{r+6} \right) \nonumber\\
&+ \left( n+ \frac12\right) \left[ \ln ( \xi r / 2)    - \ln (3 + 2r + \sqrt{3r(r+6)} \,) \right].   \label{S0-def2}
\end{align}
Here for convenience we've defined a (seemingly ubiquitous) constant 
\beq
 \xi \equiv \frac{2 + \sqrt{3}}{2 - \sqrt{3}} = (2 + \sqrt{3})^2 = 7 + 4\sqrt{3} .
\eeq

Now let us perturb about the QNM frequency, employing (\ref{om1-perturb}), to obtain
\begin{eqnarray}
S_{0n}^+(r) &=& \bar{S}_{0n}^+(r) + \sqrt{27} i \epsilon Z(r) \\
S_{0n}^-(r)  &=& \bar{S}_{0n}^+(r) - \left[ (2n+1) + \sqrt{27} i \epsilon \right] Z(r)
\end{eqnarray}
where
\begin{align}
Z^\prime(r) =& \sqrt{27} / \left[ (r-3) \sqrt{r (r+6)} \right] \\
\Rightarrow \quad Z(r) =& \ln| r-3 | \nonumber\\
&\quad - \ln \left( 3+2r + \sqrt{3r(r+6)} \, \right) + \ln \xi \, .
\end{align}
Let us now make the change of variables (\ref{z-def}) and examine close to $z = 0$:
\begin{align}
\bar{S}_0^+ &\approx \ln \left[ (4/3)^{1/4} (\xi / 12)^{n+1/2} \right], \\
Z &\approx \ln \left[  |z| / (2 \sqrt{27 L} ) \right]. \quad \quad
\end{align}
The asymptotics of the $u^{+}$ and $u^{\text{--}}$ functions defined in (\ref{upm-def}) are 
\begin{align}
u^{+}(r)  \approx& \sqrt{2} e^{+iz^2 / 2} \left(\frac{z}{L^{1/2}} \right)^n 
\nonumber\\
& \qquad\times 
\left( \frac{\xi}{4\sqrt{27}}  \right)^{N} 
 \left( \frac{\xi |z|}{ 2 \sqrt{27 L} }  \right)^{\sqrt{27} i \epsilon}   \label{uplus-asymp} \\
u^{\text{--}}(r) \approx& \sqrt{2} e^{-iz^2 / 2} \left(\frac{z}{L^{1/2}} \right)^{-(n+1)} 
\nonumber\\
& \qquad\times \left( \frac{\xi}{\sqrt{27}}  \right)^{-N}   
\left( \frac{\xi |z|}{ 2 \sqrt{27 L} }  \right)^{-\sqrt{27} i \epsilon} .  \label{uminus-asymp}
\end{align}

\subsection{Matching procedure}
Let us now define three solutions, 
\begin{eqnarray}
 u_0 &=&  D_{n + \sqrt{27} i \eps} ( (-1 + i) z) \\
 u_> &=& \Bin u^{-} + \Bout u^{+} \\
 u_< &=& \Cin u^{+}
\end{eqnarray}
Now we match $u_0$ to $u_<$ in the region $z < 0$, and match $u_0$ to $u_>$ in the region $z > 0$. Applying asymptotic results (\ref{psi-asymp1}), (\ref{psi-asymp2}), (\ref{uplus-asymp}) and (\ref{uminus-asymp}) we obtain
\begin{align}
\Cin   =& 2^{2n} 2^{(n+1)/2} L^{n/2} \left( \xi / \sqrt{27} \right)^{-N} e^{-i \pi n/4} \\
\Bout =& 2^{2n} 2^{(n+1)/2} L^{n/2} \left( \xi / \sqrt{27} \right)^{-N} e^{3 i \pi n/4} \\
=& (-1)^n \Cin \\
\Bin  =& \epsilon \, \Gamma(n+1) 2^{-n} L^{-n} \left( \frac{27 \pi}{i L} \right)^{1/2} \nonumber\\
&\qquad\times e^{-i\pi n/2} \left( \frac{\xi}{2 \sqrt{27}} \right)^{2N}
\end{align}

\subsection{Ingoing and outgoing coefficients}
In order to find the ingoing and outgoing coefficients $\Ain$ and $\Aout$, we first need the following `phase factors':
\begin{eqnarray}
\alp_1 &=& \frac{\exp \left( i \omega \int_{r=3}^{r=2} (1+6/r)^{1/2} (1-3/r) f^{-1} dr \right)}{\exp(- i \omega \rstar) } \nonumber \\
&=& \exp \left(  i \omega  [6-\sqrt{27} + 8\ln2 - 3\ln \xi ] \right) \\
\bet_1 &=&  \frac{\exp \left( i \omega \int_{r=3}^{r=\infty} (1+6/r)^{1/2} (1-3/r) f^{-1} dr \right)}{\exp(+ i \omega \rstar) }  \nonumber\\
 &=& \exp \left(  i \omega  [3-\sqrt{27} + 4\ln2 - 3 \ln \xi ] \right) \\ 
\gam_1 &=&  \frac{\exp \left( - i \omega \int_{r=3}^{r=\infty} (1+6/r)^{1/2} (1-3/r) f^{-1} dr \right)}{\exp(- i \omega \rstar) }\nonumber \\
 &=&  1/\beta_1 \\
\alp_2 &=& \lim_{r \rightarrow 2} e^{S_{0n}^+}  = 1 \\
\bet_2 &=& \lim_{r \rightarrow \infty} e^{S_{0n}^+} =  2^{1/2} \left( \xi^{1/2} / 2 \right)^{N} \\
\gam_2 &=& \lim_{r \rightarrow \infty} e^{S_{0n}^-} = 2^{1/2} \left( 2 \xi^{1/2} \right)^{-N}
\end{eqnarray}

Then, to leading order in $L$,
\begin{align}
\Aout =& \frac{\bet_1 \bet_2}{\alp_1 \alp_2} \, \frac{\Bout}{\Cin} \nonumber\\
=& 2^{1/2} (-1)^n e^{-i \omega(3+4\ln2)} \left( \xi^{1/2}/2 \right)^{N}  \label{Aout-result}
 \\
\Ain =& \frac{\gam_1 \gam_2}{\alp_1 \alp_2} \, \frac{\Bin}{\Cin} \nonumber\\
=&
 \eps \, \Gamma(n+1) 2^{-n} L^{-n} \left( \frac{27 \pi}{i L} \right)^{\!\! 1/2} \!\! e^{-i \pi n/2} \left( \frac{\xi}{2 \sqrt{27}} \right)^{\! 2 N } \nonumber \\
  & \> \times \, e^{-i \omega [9+2\sqrt{27} + 12\ln 2 - 6\ln \xi]} 2^{1/2}\left( 2 \xi^{1/2} \right)^{-N}  \label{Ain-result}
\end{align}

\subsection{Excitation factors}
The excitation factors $\Bef$ are defined in Eq.~(\ref{Bef-def}). Combining results (\ref{Ain-result}) and (\ref{Aout-result}) we obtain the leading-order approximation
\begin{align}
\mathcal{B}_{ln} \approx \frac{(-i L)^{n-1/2}}{n! \, (\omega_{ln} / [\sqrt{27} L])} \frac{\exp(2i \omega_{ln} \zeta)}{\sqrt{8 \pi}} \left( \frac{216}{\xi}  \right)^{N} 
\end{align}
with error $\mathcal{O}(L^{n-3/2}) $ and where the constant $\zeta$ is defined in (\ref{y-def}). This can be rewritten as Eq.~(\ref{Bef-approx}), the result given in the main text.

Expression (\ref{Bef-approx}) fits the large-$l$ numerical data \cite{Berti-Cardoso-2006} well, for $n=0, 1, 2$. 
It is considerably more effort to obtain the higher order corrections and we shall present  these  together with efficient numerical techniques for their calculation in a subsequent publication.

\bibliographystyle{apsrev}


\end{document}